\documentclass[twocolumn,pre,showpacs,amsmath,amssymb]{revtex4}
\usepackage{graphicx}% Include figure files
\usepackage{dcolumn}% Align table columns on decimal point
\usepackage{bm}
\usepackage{color}

\usepackage{amssymb}
\newcommand{\NS}{\text{NS}}

\newcommand{\ltu}{\text{th}}

\newcommand{\dd}{\text{d}}

\newcommand\beq{\begin{equation}}
\newcommand\eeq{\end{equation}}
\newcommand\beqa{\begin{eqnarray}}
\newcommand\eeqa{\end{eqnarray}}
\newcommand{\nn}{\nonumber\\}

\newcommand{\ed}{\end{document}}

\begin{document}
\title{{Class of dilute granular Couette flows with uniform heat flux}}
\author{Francisco Vega Reyes}
\author{Vicente Garz\'o}
\author{Andr\'es Santos}
\affiliation{Departamento de F\'{\i}sica, Universidad de Extremadura, E-06071
Badajoz, Spain}

\begin{abstract}
In a recent paper [F. Vega Reyes \emph{et al.}, Phys. Rev. Lett. \textbf{104}, 028001 (2010)]
we presented a preliminary description of a special class of steady Couette flows in dilute granular gases. In all flows of this class the viscous heating is exactly balanced by inelastic cooling. This yields  a uniform heat flux and a linear relationship between the local temperature and flow velocity. The class (referred to as the LTu class) includes the Fourier flow of ordinary gases and the simple shear flow of granular gases as special cases. In the present paper we provide further support for this class of Couette flows by following four different routes, two of them being theoretical (Grad's moment method of the Boltzmann equation and exact solution of a kinetic model) and the other two being computational (molecular dynamics and Monte Carlo simulations of the Boltzmann equation). Comparison between theory and simulations shows a very good agreement for the non-Newtonian rheological properties, even for quite strong inelasticity, and a good agreement for the heat flux coefficients in the case of Grad's method, the agreement being only qualitative in the case of the kinetic model.
\end{abstract}


\date{\today}
\maketitle

\section{Introduction}
\label{sec1}

The development of the kinetic theory of non-uniform gases, extending the results by Boltzmann \cite{B95} and Maxwell \cite{M67} to near-equilibrium systems, started out with the seminal works, in the early 20th Century, by Hilbert \cite{Hilbert}, Enskog \cite{B66}, Chapman \cite{CC70}, and Burnett \cite{B34}. Their results allow for an accurate description of non-equilibrium states of gases (in particular, neutral gases), in the limit of Newtonian hydrodynamics \cite{CC70} (that is, small gradients, scaled with the typical microscopic  length scale, of the average fields). These theoretical works have been  recently extended to the more general frame of \textit{granular} gases where the inter-particle collisions are inelastic \cite{H83,JNB96a,BP04}. The prototypical model of a granular fluid consists of a system of smooth inelastic hard spheres with a constant coefficient of restitution $\alpha$. This parameter distinguishes ordinary gases ($\alpha=1$) from granular gases ($\alpha<1$).

Granular matter is certainly involved, not only in many industrial processes \cite{M93}, but also {in} biological processes \cite{AT06,UMS96}. {This explains the  {growing} interest in the study of granular matter}. Moreover, granular flows are also challenging from a more fundamental point of view \cite{C90,Go03}.
For instance, in the low-density regime, the Boltzmann equation can be generalized to granular gases. For {all} these reasons there is currently a great interest in the study of granular matter and a large number of research works have been recently published in this field {(see, for instance, Refs.\ \cite{AT06,BP04,JNB96a,K99,OK00,D00,K04} and references therein)}. In particular, the Navier--Stokes (NS) constitutive hydrodynamic equations for granular gases have been derived from the Boltzmann and Enskog equations \cite{JM89,GS95,BDKS98,GD99,BC01,GD02,L05,GDH07,GHD07}. This has allowed {the description of} important phenomena in granular matter, some of which  were found to persist with the same qualitative behavior even beyond the range of Newtonian hydrodynamics (basic segregation mechanisms \cite{K04}, for instance).

Unfortunately, the ranges of interest of the physics of granular gases {fall} frequently beyond Newtonian hydrodynamics since the strength of the spatial gradients is large in most situations of practical interest (for example, in steady states), due to the coupling between inelasticity and gradients \cite{Go03,SGD04}. In these states, a hydrodynamic description is still valid but with constitutive equations more complex than the  NS ones.
On the other hand, the derivation of these non-Newtonian equations from the inelastic Boltzmann equation is an extremely complex mathematical task. For this reason, one is forced to resort to approximate schemes (such as Grad's 13-moment method or the use of simplified kinetic models), to be tested against computer simulations such as the direct simulation Monte Carlo (DSMC) method \cite{B94} and {event-driven} molecular dynamics (MD) {simulations} \cite{R04}. In this context,  analytical solutions of the Bhatnagar--Gross--Krook (BGK) model kinetic equation, and its extension to inelastic collisions, have been found for steady non-linear shear flows,  both for {elastic  \cite{GS03} and  granular gases \cite{BRM97,TTMGSD01,L06,G06}.} Comparison with numerical solutions of the Boltzmann equation by means of the DSMC method shows that this kinetic model is able to describe the general properties of  non-linear shear flows in elastic and granular gases.

One of the well-known examples of steady states is the simple or uniform shear flow (USF) problem \cite{Go03,SGD04}. This state is characterized by a linear velocity field (that is, $\partial u_x/\partial y=\text{const}$), constant density $n$, and constant temperature $T$. The presence of shearing induces anisotropies in the pressure tensor $P_{ij}$, namely, nonzero shear stress $P_{xy}$ and normal stress differences $P_{xx}-P_{yy}$ and $P_{yy}-P_{zz}$. On the other hand, the heat flux vanishes due to the absence of density and thermal gradients. The steady-state condition requires that the collisional cooling (which is fixed by the mechanical properties of the {granular gas particles}) is exactly balanced by viscous heating {(which is fixed by the shearing)}. This relationship between the shear field and dissipation sets the  strength of the \emph{scaled} velocity gradient  for a given value of the coefficient of restitution.
This implies that the corresponding hydrodynamic steady state is inherently non-Newtonian ({that is,} beyond the scope of the NS equations) in  inelastic granular gases \cite{SGD04}.

\begin{figure}
\includegraphics[width=0.95\columnwidth]{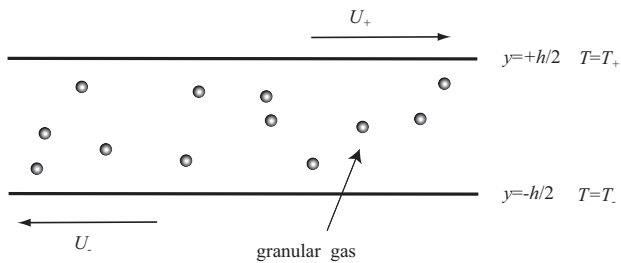}
\caption{The planar Couette flow is driven by two horizontal plates, separated by a distance
$h$. Both act like sources of temperature and shear on a {low-density} granular gas filling the space between them.} \label{fig1}
\end{figure}

Let us  consider the more complex case of a generic planar Couette flow problem{, which is depicted in Fig.\ \ref{fig1}}. In this state, the temperature is {in principle} not uniform and, consequently, a heat flux vector $\mathbf{q}$ coexists with the pressure tensor $P_{ij}$ \cite{TTMGSD01}. In fact, the energy balance equation (in the steady state) reads
\begin{equation}
\frac{\partial q_y}{\partial y}=-\frac{d}{2}\zeta nT-P_{xy}\frac{\partial u_x}{\partial y},
\label{Tbal}
\end{equation}
where  $\zeta$ is the inelastic cooling rate and $d$ is the dimensionality of the system. The first term on the right-hand side is an energy sink term reflecting the dissipation due to collisions, while the second term [note that $\text{sgn}(P_{xy})=-\text{sgn}(\partial u_x/{\partial y})$] is an energy source term due to viscous heating. The competition between these two terms determines the sign of the divergence of the heat flux \cite{VU09}. As for the conservation equation for momentum, it  implies
\beq
P_{xy}=\text{const},
\label{Pxy}
\eeq
\beq
P_{yy}=\text{const}.
\label{Pyy}
\eeq

In general, Eq.\ \eqref{Tbal} applies to any state that (i) is stationary, (ii) has gradients only along the $y$ direction, and (iii) has a flow velocity vector along the $x$ direction. Thus, Eq.\ \eqref{Tbal} is also valid for the familiar Fourier flow of ordinary gases ($\alpha=1$) {as well as} for the (steady-state) USF of granular gases ($\alpha<1$). In the first case $\zeta=0$ and  $\partial u_x/{\partial y}=0$, so the non-zero heat flux vector is uniform. In the second case, there is no heat flux and, as said before, the condition
\beq
\zeta =-\frac{2}{d}\frac{P_{xy}}{nT}\frac{\partial u_x}{\partial y}
\label{Tbal2}
\eeq
establishes the relationship between the inelastic cooling and the shear field.
These two clearly distinct states share the common features of uniform heat flux and a local balance between inelastic cooling and viscous heating. The interesting question is, does there exist a whole class of Couette flows also sharing the same features? This class would include the Fourier flow of elastic gases and the USF of inelastic gases as special limit situations.

The aim of this paper is to provide numerical and analytical evidence on the existence of such a class of Couette flows. On the numerical side, we have solved the inelastic Boltzmann equation by means of the {DSMC} method \cite{B94} and have carried out {MD} simulations of dilute granular gases. On the analytical side, we have solved this special class of Couette flows from a simplified model kinetic equation as well as by the application of Grad's 13-moment method to the Boltzmann equation. A further theoretical support for this class has recently been found from an exact solution of the Boltzmann equation for inelastic Maxwell models \cite{SGV09}.
Apart from the condition $\mathbf{q}=\text{const}$, this  class of Couette flows is macroscopically characterized by a uniform pressure,
\beq
p=nT=\text{const},
\label{p}
\eeq
and
\begin{equation}
\nu^{-1}\partial_y T=A=\text{const},
\label{1bis}
\end{equation}
\begin{equation}
\nu^{-1}\partial_y u_x=a(\alpha)=\text{const},
\label{1}
\end{equation}
where
$\nu\propto n T^{1/2}$ is an effective (local) collision frequency. As a consequence of Eqs.\ \eqref{1bis} and \eqref{1}, while neither $u_x(y)$ nor $T(y)$ are linear, a parametric plot of $T$ vs $u_x$ shows a \emph{linear} relationship. For this reason, we refer to this class of flows as ``linear $T(u_x)$'' flows, or simply, ``LTu'' flows. The slope of the linear plot $T(u_x)$ goes from zero in the inelastic USF limit (constant temperature) to infinity in the elastic Fourier flow (zero macroscopic velocity).
As we will see, the transport properties in the LTu class are highly non-Newtonian  and can be characterized by a generalized shear viscosity, normal stress differences, a generalized thermal conductivity, and a cross coefficient associated with the $x$ component of the heat flux.
A preliminary  report of the LTu has been published recently \cite{VSG10}.

The paper is organized as follows. In Sec.\ \ref{sec2} we present the formal description at a kinetic theory level of the LTu flows, derive the relation between the Reynolds number and the relevant parameters, and define the generalized transport coefficients. We find in Sec.\ \ref{sec3} two analytical solutions of the problem introduced: in Sec.\  \ref{Grad} an approximate analytical solution  is obtained by means of {Grad's} 13-moment method, whereas in Sec.\  \ref{BGK} we find an exact solution of a model kinetic equation (BGK-type kinetic model adapted to the granular gas {\cite{BDS99}}). In Sec.\  \ref{sec4} the simulation techniques (both DSMC and MD) used in this work are described. Theory and simulation results are shown and compared in Sec.\ \ref{sec5}. {Finally,} in Sec.\  \ref{sec6} we give a brief summary of results and discuss them.

\section{Boltzmann description of the LTu flow}
\label{sec2}
\subsection{Couette flow}
Let us consider a granular fluid modeled as a gas of inelastic hard spheres.
A constant parameter, the coefficient of normal restitution  $\alpha$, accounts for the inelasticity {in collisions}. Its values range from $\alpha=0$ (purely inelastic collision) to $\alpha=1$ (purely elastic collision).
In the low-density regime,
the one-particle velocity distribution function $f(\mathbf{r},\mathbf{v};t)$ obeys the inelastic Boltzmann equation \cite{BP04,GS95}
\beq
\left(\partial_t
+\mathbf{v}\cdot\nabla\right)f(\mathbf{r},\mathbf{v};t)=J[\mathbf{v}|f,f],
\label{2.1}
\eeq
where the Boltzmann collision operator $J[\mathbf{v}|f,f]$ is given
by
\beqa
J\left[{\bf v}_{1}|f,f\right]& =&\sigma^{d-1} \int
\dd{\bf v}_{2}\int \dd\widehat{\boldsymbol{\sigma}} \, \Theta\left(\mathbf{g}\cdot \widehat{\boldsymbol{\sigma}}\right)\left(\mathbf{g}\cdot \widehat{\boldsymbol{\sigma}}\right)\nn
&&\times\left[
\alpha^{-2}f({\bf v}_{1}')f({\bf v}_{2}')-f({\bf v}_{1})f({\bf
v}_{2})\right] .
\label{2.2}
\eeqa
Here, $\sigma$ is the diameter of a sphere, $\Theta(x)$ is Heaviside's step function, $\widehat{\boldsymbol{\sigma}}$ is a unit vector
directed along the centers of the two colliding particles, $\mathbf{g}=\mathbf{v}_1-\mathbf{v_2}$ is the relative velocity, and  the primes on the velocities denote the
initial values $\{{\bf v}_{1}^{\prime}, {\bf v}_{2}^{\prime}\}$ that
lead to $\{{\bf v}_{1},{\bf v}_{2}\}$ following a binary collision:
\beqa
\label{2.3b}
{\bf v}_{1}^{\prime}&=&{\bf v}_{1}-\frac{1}{2}\left( 1+\alpha
^{-1}\right)(\widehat{\boldsymbol{\sigma}}\cdot {\bf
g})\widehat{\boldsymbol {\sigma}}, \nn
{\bf v}_{2}^{\prime}&=&{\bf
v}_{2}+\frac{1}{2}\left( 1+\alpha^{-1}\right)
(\widehat{\boldsymbol{\sigma}}\cdot {\bf
g})\widehat{\boldsymbol{\sigma}}.
\eeqa

At a hydrodynamic level, the relevant quantities are the number density $n$, the flow velocity $\mathbf{u}$, and the granular temperature $T$. They are defined as moments of the velocity distribution as
\beq
n=\int\dd \mathbf{v}\, f(\mathbf{v}),
\eeq
\beq
\mathbf{u}=\frac{1}{n}\int\dd \mathbf{v} \,\mathbf{v}f(\mathbf{v}),
\eeq
\beq
T=\frac{m}{dn}\int\dd \mathbf{v}\, V^2 f(\mathbf{v}),
\label{granT}
\eeq
where $m$ is the mass of a particle and $\mathbf{V}=\mathbf{v}-\mathbf{u}(\mathbf{r})$ is the peculiar
velocity.

The Boltzmann collision operator conserves the number of particles and the momentum, but the kinetic energy is not conserved. The corresponding balance equations are obtained by multiplying both sides of Eq.\ \eqref{2.1} by $1$, $\mathbf{v}$, $v^2$, and integrating over velocity. The result is
\beq
D_t n+n\nabla\cdot \mathbf{u}=0,
\label{n1.3}
\eeq
\beq
D_t\mathbf{u}+\frac{1}{mn}\nabla\cdot\mathsf{P}=\mathbf{0},
\label{n1.4}
\eeq
\beq
D_tT+\frac{2}{dn}\left(\nabla\cdot\mathbf{q}+\mathsf{P}:\nabla
\mathbf{u}\right)=-\zeta T.
\label{n1.5}
\eeq
Here, $D_t\equiv\partial_t+\mathbf{u}\cdot\nabla$ is the material
time derivative,
\beq
P_{ij}=m\int\dd \mathbf{v} \,V_i V_j f(\mathbf{v})
\label{Pij}
\eeq
is the pressure tensor,
\beq
\mathbf{q}=\frac{m}{2}\int\dd \mathbf{v}\,
V^2\mathbf{V}f(\mathbf{v}),
\label{qi}
\eeq
is the heat flux, and
\beq
\zeta=-\frac{m}{dnT}\int\dd \mathbf{v}\, V^2 J[\mathbf{v}|f,f]
\label{zeta}
\eeq
is the cooling rate characterizing the rate of energy dissipated due to collisions.

In the planar Couette flow the
granular gas is enclosed between two parallel, infinite plates (normal to the $y$ axis) at $y=\pm h/2$ in relative motion along the $x$ direction, and kept, in general, at different temperatures (cf.\ Fig.\ \ref{fig1}). The resulting flow velocity is along the $x$ axis and, from symmetry, it is expected that the hydrodynamic fields only vary in the $y$ direction. Consequently,  the velocity distribution function is also assumed to depend  on the coordinate $y$ only. Moreover, we focus on the steady state, so Eq.\ \eqref{2.1} becomes
\beq
v_y\partial_y f=J[\mathbf{v}|f,f].
\label{2.3}
\eeq
Under the above conditions, the mass conservation equation \eqref{n1.3} is identically satisfied, the momentum conservation equation \eqref{n1.4} reduces to $\partial_y P_{iy}=0$ {[cf.\ Eqs.\ \eqref{Pxy} and \eqref{Pyy}]}, while the energy balance equation \eqref{n1.5} becomes Eq.\ \eqref{Tbal}.
It must be noted that {Eqs.\ \eqref{Tbal}--\eqref{Pyy}} are exact consequences of the geometry of the problem and the steady-state condition. Therefore, they are valid whether a hydrodynamic description applies or not, even near the walls where boundary effects are not negligible.

Now we assume that the separation $h$ between the walls is large enough (that is, it comprises a sufficient number of mean free paths) as to identify a \emph{bulk} region where a hydrodynamic description is expected to apply.  Here the term ``hydrodynamics'' is employed in a wide sense encompassing both Newtonian and non-Newtonian behavior. In the context of the Boltzmann equation, a hydrodynamic description is linked to  a \emph{normal} solution, {namely,} a special solution where all the space and time dependence of the velocity distribution function takes place via a functional dependence on the hydrodynamic fields \cite{BDKS98}:
\beq
f=f[\mathbf{v}|n,\mathbf{u},T].
\label{normal}
\eeq

\subsection{LTu flow}
In the general Couette flow problem, the imposed velocity and temperature gradients can be controlled independently of the coefficient of restitution via the boundary conditions. This problem was studied by means of a simple kinetic model in Ref.\ \cite{TTMGSD01}. Here, however, as said in the Introduction,  we focus on a special class of Couette flows. More specifically, we assume that there exists a normal solution of the Boltzmann equation \eqref{2.3} with a uniform heat flux component $q_y$. As a consequence, the shear rate {$\partial u_x/\partial y$} is not a free parameter but it is fixed by the value of the coefficient of restitution [cf.\ Eq.\ \eqref{Tbal2}].

As indicated by Eq.\ \eqref{normal}, {we need to specify the form of the hydrodynamic fields in order to} characterize the normal solution corresponding to the class of Couette flows with uniform heat flux. This is a non-trivial risky task since the proposed spatial dependence of the fields must be consistent with Eqs.\ \eqref{Pxy} and \eqref{Pyy} and, moreover, the state is expected to lie outside the realm of the NS regime.

We take two basic assumptions {(which have already been shown to be fulfilled for generic Couette granular flows \cite{TTMGSD01})}. First, the exact condition \eqref{Pyy} is extended to the remaining diagonal elements of the pressure tensor, so that its trace is also uniform. This first assumption is displayed in Eq.\ \eqref{p}. Note that in the NS description, $p=P_{yy}$, so Eq.\ \eqref{p} is a straightforward consequence of the conservation of momentum. Here, however, we assume Eq.\ \eqref{p} even though, as will be seen below, $p\neq P_{yy}$. The second assumption is subtler and refers to the $y$ component of the heat flux. According to the concept of a normal solution $q_y=q_y[n,\mathbf{u},T]$ is a functional of the hydrodynamic fields. We assume that such a functional dependence has the same form as in the NS description, namely, $q_y\propto (p/nT^{1/2})\partial_y T$. Note, however, that the proportionality constant is in general different from the NS one. Since $p$ has already been assumed to be uniform and $q_y=\text{const}$ defines the LTu state, it follows Eq.\ \eqref{1bis} with $\nu\propto n T^{1/2}$. Therefore, Eqs.\ \eqref{p} and \eqref{1bis} define the assumed hydrodynamic profiles. The energy balance equation \eqref{Tbal2} yields Eq.\ \eqref{1}, where we take into account that $\zeta\propto \nu$ as well as Eqs.\ \eqref{Pxy} and \eqref{p}.
The constant $a(\alpha)$ is a dimensionless parameter that plays the role of the Knudsen number {($\text{Kn}$)} associated with the shearing. As indicated by the notation,  $a(\alpha)$ is not a free parameter but depends on the coefficient of restitution through Eq.\ \eqref{Tbal2}. On the other hand, the constant parameter $A$ defined by Eq.\ \eqref{1bis} is not constrained by the value of $\alpha$. Note that $A$ is not a dimensionless number, the corresponding Knudsen number associated with the thermal gradient being {$\epsilon\equiv A/\sqrt{mT}$.}

As said in Sec.\ \ref{sec1}, from Eqs.\ \eqref{1bis} and \eqref{1} one obtains
\beq
\frac{\partial T}{\partial u_x}=\frac{A}{a(\alpha)}=\text{const}.
\label{LTu}
\eeq
This means that if the  spatial coordinate $y$ normal to the moving plates is eliminated between temperature and flow velocity the resulting profile $T(u_x)$ is  \emph{linear}, thus justifying the acronym LTu used here to refer to this class of flows.

It is interesting to  get the explicit spatial dependence of $T$ and $u_x$ \cite{VU09}. From Eq.\ \eqref{1bis} it is easy to obtain
\beq
T(y)=T_0\left[1+\frac{3A\nu_0}{2T_0}(y-y_0)\right]^{2/3},
\label{21}
\eeq
where $y_0$ is an arbitrary reference point in the bulk region, and $T_0$ and $\nu_0$ are the local values of $T$ and $\nu$, respectively, at $y=y_0$. Integrating Eq.\ \eqref{LTu}, with the aid of  Eq.\ \eqref{21}, we simply get
\beq
u_x(y)=\frac{a(\alpha)}{A}T_0\left[1+\frac{3A\nu_0}{2T_0}(y-y_0)\right]^{2/3}+u_0-\frac{a(\alpha)}{A}T_0,
\label{2.1x}
\eeq
where $u_0$ is the local value of $u_x$ at $y=y_0$.
{The expression of the (local) thermal Knudsen number is
\beq
\epsilon(y)=\frac{A}{\sqrt{mT_0}}\left[1+\frac{3A\nu_0}{2T_0}(y-y_0)\right]^{-1/3}.
\label{epsilon}
\eeq
}

\begin{figure}
\includegraphics[width=0.95\columnwidth]{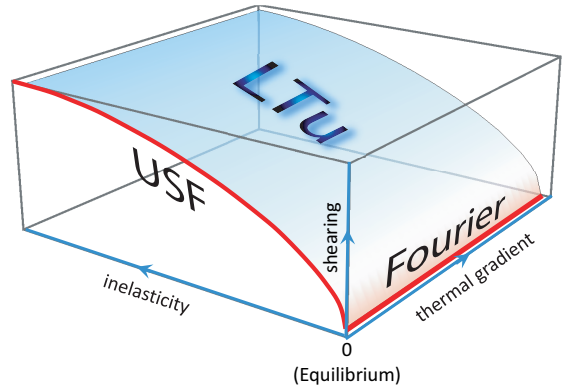}
\caption{(Color online) Each point of this diagram represents a  Couette flow steady state. The  surface defines the LTu class, which contains the lines representing the Fourier flow for ordinary gases {(that is, no shearing and no inelasticity)} and the USF for granular gases {(that is, no thermal gradient)}.\label{fig2}}
\end{figure}

In the particular case of elastic particles ($\alpha=1$ or, equivalently, $\zeta=0$), Eq.\ \eqref{Tbal2} implies $a=0$, so we recover the Fourier flow of ordinary gases \cite{MMSG94}. On the other hand, in the absence of thermal gradients ($A\to 0$) but in the presence of inelastic collisions {($\alpha<1$)}, Eqs.\ \eqref{21} and \eqref{2.1x} become $T=T_0$ and $u_x=u_0+a(\alpha)\nu_0(y-y_0)$, that is, we recover the conditions of USF. For general values of $\alpha$ and $A$, Eqs.\ \eqref{p}--\eqref{1} define a whole class of Couette flows with uniform $q_y$. This manifold of Couette states is sketched in Fig.\ {\ref{fig2}}.
On the LTu surface one has $\partial_y q_y=0$, while the points above (below) the surface represent Couette-flow states where the dominant term in Eq.\ \eqref{Tbal} is the viscous heating (inelastic cooling) one and thus $\partial_y q_y>0$ ($\partial_y q_y<0$). For an analysis of the curvature of the temperature profiles within the NS domain, see Ref.\ \cite{VU09}.

\subsection{Reynolds number for LTu flows\label{Rey}}

So far, we have not needed to specify the explicit form of the effective collision frequency $\nu$, except for the scaling relation $\nu\propto n T^{1/2}$. Henceforth, we will adopt for $\nu$ the conventional choice of effective collision frequency in shear flow problems {involving ordinary gases}, namely,
\beq
\nu=\frac{p}{\eta_\NS^0},
\label{etaNS0}
\eeq
where $\eta_\NS^0$ is the NS shear viscosity of a gas of elastic hard spheres. With this choice, one has (in the leading Sonine approximation) \cite{CC70}
\begin{equation}
\label{3.5}
\nu=\frac{8\pi^{(d-1)/2}}{(d+2)\Gamma\left(\frac{d}{2}\right)} n\sqrt{\frac{T}{m}}\sigma^{d-1}.
\end{equation}

It is instructive to express the Reynolds number of the LTu flow in terms of the reduced shear rate $a(\alpha)$, the wall temperatures $T_\pm$, the slab width $h$, and a nominal mean free path $\bar{\ell}$.
The Reynolds number $\text{Re}$ is defined as \cite{T88}
\beq
\text{Re}=\frac{m\bar{n}(U_+-U_-)h}{\bar{\eta}_\NS^0},
\label{R1}
\eeq
where $\bar{n}$ and $\bar{\eta}_\NS^0$ are characteristic values for density and shear viscosity, respectively. Here we take $\bar{n}$ as the average number density and   $\bar{\eta}_\NS^0=p/\bar{\nu}$, where $\bar{\nu}$ is given by Eq.\ \eqref{3.5} by setting $n=\bar{n}$ and $T=T_-$.

Neglecting velocity slips and temperature jumps near the walls, and choosing $y_0=-h/2$ in Eqs.\ \eqref{21} and \eqref{2.1x}, one obtains
\beqa
U_+-U_-&=&\frac{a(\alpha)}{A}\left(T_+-T_-\right)\nn
&=&\frac{3}{2}\frac{\Delta T}{(1+\Delta T)^{3/2}-1}a(\alpha)\nu(-h/2)h,
\label{R2}
\eeqa
where $\Delta T\equiv T_+/T_- -1$ and, without loss of generality, we have assumed $T_+\geq T_-$.
Insertion of Eq.\ \eqref{R2} into Eq.\ \eqref{R1} yields
\beq
\text{Re}=\frac{3}{2}\frac{\Delta T}{(1+\Delta T)^{3/2}-1}a(\alpha)\left(\frac{h}{\bar{\ell}}\right)^2,
\label{R3}
\eeq
where $\bar\ell\equiv \sqrt{T_-/m}/\bar{\nu}$ is the nominal mean free path. Upon derivation of Eq.\ \eqref{R3} use has been made of the relation $\nu(-h/2)/\bar{\nu}=n(-h/2)/\bar{n}=p/\bar{n}T_-$.
Equation \eqref{R3} expresses the Reynolds number in terms of the relative temperature difference $\Delta T$, the shear-rate Knudsen number $a(\alpha)$, and the system-size Knudsen number $\bar\ell/h$. We observe that $\text{Re}$ is essentially the ratio between the shear-rate Knudsen number and the square of the system-size Knudsen number. The pre-factor depends on $\Delta T$ and ranges from $1$ in the limit $\Delta T\to 0$ to $0$ in the opposite limit $\Delta T\to\infty$.

\subsection{Non-Newtonian transport coefficients}
As said above, the LTu flow is in general non-Newtonian. This can be characterized by {the introduction of} \emph{generalized} transport coefficients measuring the relationship between momentum and heat fluxes with the hydrodynamic gradients. First, we define a non-Newtonian shear viscosity coefficient $\eta(\alpha)$ by
\beq
P_{xy}=-\eta(\alpha)\frac{\partial u_x}{\partial u_y}.
\label{eta}
\eeq
Since, by dimensional analysis, $\eta\propto p/\nu$, Eq.\ \eqref{eta} is consistent with Eqs.\ \eqref{Pxy}, \eqref{p}, and \eqref{1}. Equation \eqref{eta} can be seen as a generalization of the NS constitutive equation for the shear stress in the sense that it is assumed that $P_{xy}$ is independent of the thermal gradient {$A$}. On the other hand, the generalized shear viscosity coefficient $\eta(\alpha)$ is expected to differ from the NS {shear viscosity coefficient} $\eta_\NS(\alpha)$ {of an inelastic dilute gas \cite{BDKS98}}.
{The energy balance equation \eqref{Tbal2} establishes a relationship between the reduced shear rate $a(\alpha)$, the generalized shear viscosity $\eta(\alpha)$, and the cooling rate $\zeta(\alpha)$:}
\beq
a^2(\alpha)=\frac{d}{2}\frac{\zeta^*(\alpha)}{\eta^*(\alpha)},
\label{3.4}
\eeq
where $\zeta^*\equiv \zeta/\nu$ and $\eta^*\equiv \eta/(p/\nu)$.

{While $P_{xx}=P_{yy}=p$ in the NS regime, normal stress differences are expected to appear.} They can be measured though the coefficients
\beq
\frac{P_{xx}}{p}=\theta_x(\alpha),\quad \frac{P_{yy}}{p}=\theta_y(\alpha).
\label{theta}
\eeq
{For $d\geq 3$, one} could define a coefficient $\theta_z={P_{zz}/p}$ but it is related to $\theta_x$ and $\theta_y$ by the condition $\theta_x+\theta_y+(d-2)\theta_z=d$. The quantities $\theta_x$ and $\theta_y$ {represent}  \emph{directional} temperatures $T_x=P_{xx}/n$ and  $T_y=P_{yy}/n$ (relative to the granular temperature $T$) along the $x$  and $y$ directions, respectively.

In the case of the heat flux, the assumed scaling relation $q_y\propto (p/\nu)\partial_y T$ suggests the Introduction of a generalized thermal conductivity coefficient $\lambda(\alpha)$ as
\beq
q_y=-\lambda(\alpha)\frac{\partial T}{\partial y}.
\label{lambda}
\eeq
This equation has the same form as Fourier's law, except that the coefficient $\lambda(\alpha)$ is expected to differ from the corresponding NS thermal conductivity coefficient {of an inelastic dilute gas \cite{BDKS98}}.
Moreover, while $q_x=0$ in the NS description, here we assume the existence of a nonzero $x$ component of the heat flux due to a non-Newtonian coupling between shearing and temperature gradient. To characterize this non-Newtonian effect,  we introduce a cross coefficient $\phi(\alpha)$ as
\beq
q_x=\phi(\alpha)\frac{\partial T}{\partial y}.
\label{phi}
\eeq
Dimensional analysis shows that {$\lambda\propto p/\nu$ and $\phi\propto p/\nu$, so that Eqs.\ \eqref{lambda} and \eqref{phi} imply that $\mathbf{q}$ is uniform.}

It must be borne in mind that in this section we have \emph{assumed} the existence of Couette flows with (a) $q_y=\text{const}$ and (b) profiles given by Eqs.\ \eqref{p}--\eqref{1}, but there is no {\emph{a priori}} guarantee that the Boltzmann equation \eqref{2.3} admits such states. In the next  section we will provide support for the existence of this LTu class by solving Eq.\ \eqref{2.3} through the approximate Grad 13-moment method and by an exact solution of a model kinetic equation of the {inelastic} Boltzmann equation. Further support will be given by computer simulations, showing a good agreement with some of the theoretical results.

\section{Theoretical approaches}
\label{sec3}

\subsection{Grad's moment method}
\label{Grad}
In order to check the consistency of the hydrodynamic profiles \eqref{p}--\eqref{1}, as well as of the momentum and heat fluxes, here we will solve the Boltzmann equation by the classical Grad moment method \cite{G49}. This in turn will provide explicit expressions for the generalized transport coefficients $\eta$, $\theta_i$, $\lambda$, and $\phi$.

The idea {behind} Grad's moment method is to expand the velocity distribution function $f$ in a complete set of orthogonal polynomials (generalized Hermite polynomials), the coefficients being the corresponding velocity moments. Next, the expansion is truncated after a certain order $k$. When this truncated expansion is substituted into the hierarchy of moment equations up to order $k$ one gets a closed set of coupled equations. In the standard 13-moment method the retained moments {are} the hydrodynamic fields ($n$, $\mathbf{u}$, and $T$) plus the irreversible momentum and heat fluxes ($P_{ij}-p\delta_{ij}$ and $\mathbf{q}$). More explicitly,
\beqa
\label{3.1}
f&\to& f_0\Bigg\{1+\frac{m}{2nT^2}\Big[\left(P_{ij}-p\delta_{ij}\right)V_iV_j\nn
&&+\frac{4}{d+2}\left(
\frac{mV^2}{2T}-\frac{d+2}{2}\right){\bf V}\cdot {\bf q}\Big]\Bigg\},
\eeqa
where
\beq
f_0=n\left(\frac{m}{2\pi T}\right)^{d/2} e^{-mV^2/2T}
\label{3.2}
\eeq
is the local equilibrium distribution.
In the three-dimensional case, there are 13 moments involved in Eq.\ \eqref{3.1}; hence  this method is referred to as the 13-moment method. In the case of {a general dimensionality} $d$ the number of moments is $d(d+5)/2+1$.

In order to have a closed set of equations for $n$, $\mathbf{u}$, $T$, $P_{ij}-p\delta_{ij}$, and $\mathbf{q}$ we need to make use of Eq.\ \eqref{3.1} to get
\beq
\frac{m}{2}\int\dd\mathbf{v}\,V_iV_jV_k f\to\frac{1}{d+2}\left(q_i\delta_{jk}+q_j\delta_{ik}+q_k\delta_{ij}\right),
\label{G1}
\eeq
\beq
\frac{m}{2}\int\dd\mathbf{v}\,V^2 V_i V_j f\to \frac{p}{nm}\left(\frac{d+4}{2} P_{ij}-p\delta_{ij}\right).
\label{G2}
\eeq
Moreover, the collisional moments associated with the momentum and energy transfers are approximated by
\begin{equation}
\label{3.3}
m\int \dd{\bf V}\,V_iV_jJ[f,f]\to-\beta_1\nu \left(P_{ij}-p\delta_{ij}\right)-\zeta P_{ij},
\end{equation}
\begin{equation}
\label{3.4x}
\frac{m}{2}\int \dd{\bf V}\,V^2{\bf V}J[f,f]\to-\frac{d-1}{d}\beta_2\nu {\bf q},
\end{equation}
where
\beq
\zeta=\nu\frac{d+2}{4d}(1-\alpha^2),
\label{3.7}
\eeq
\begin{equation}
\label{3.6}
\beta_1=\frac{1+\alpha}{2}\left[1-\frac{d-1}{2d}(1-\alpha)\right],
\end{equation}
\begin{equation}
\label{3.8}
\beta_2=\frac{16+11d-3(d+8)\alpha}{16(d-1)}(1+\alpha).
\end{equation}
It is important to remark that, upon writing Eqs.\ \eqref{3.3} and \eqref{3.4x}, nonlinear terms in $P_{ij}-p\delta_{ij}$ and $\mathbf{q}$ are neglected. This is the usual implementation of Grad's method, although the quadratic  terms are sometimes retained \cite{HH82,TK95}.  Note that the expression of the cooling rate $\zeta$ provided by Grad's method and given by Eq.\ \eqref{3.7} coincides with its local-equilibrium form.
The dimensionless parameters $\beta_1$ and $\beta_2$ measure the impact of inelasticity on the collisional transfer {of} momentum and energy, respectively. Both coefficients reduce to unity in the elastic limit.

Now, let us apply Grad's method to the Boltzmann equation \eqref{2.3}. In the geometry of the Couette flow, the relevant moments are $n$, $u_x$, $T$, $P_{xy}$, $P_{xx}$, $P_{yy}$, $q_x$, and $q_y$. Of course, the exact balance equations {\eqref{Tbal}--\eqref{Pyy}} are recovered. The remaining five equations are obtained by multiplying both sides of Eq.\ \eqref{2.3} by $V_x V_y$, $V_x^2$, $V_y^2$, $V^2V_x$, and $V^2V_y$, integrating over velocity, and applying the approximations \eqref{G1}--\eqref{3.4x}. The results are
\beq
\frac{2}{d+2}\partial_y  q_x+P_{yy}{\partial_y u_x}
=-\left(\beta_1\nu+\zeta\right)P_{xy},
\label{3.9}
\eeq
\beq
\frac{2}{d+2}\partial_y  q_y+2P_{xy}{\partial_y u_x}
=-\beta_1\nu\left(P_{xx}-p\right)-\zeta P_{xx},
\label{3.10}
\eeq
\beq
\frac{6}{d+2}\partial_y  q_y
=-\beta_1\nu\left(P_{yy}-p\right)-\zeta P_{yy},
\label{3.11}
\eeq
\begin{equation}
\label{3.12}
\frac{d+4}{2}\partial_y\left(\frac{T}{m}P_{xy}\right)+\frac{d+4}{d+2} q_y\partial_y u_x=-\frac{d-1}{d}\beta_2\nu
q_x,
\end{equation}
\begin{equation}
\label{3.13}
\partial_y\left[\frac{T}{m}\left(\frac{d+4}{2}P_{yy}-p\right)\right]+\frac{2}{d+2}
q_x\partial_y u_x=-\frac{d-1}{d}\beta_2\nu q_y,
\end{equation}

We have made no {extra} assumptions in the set of equations  \eqref{3.9}--\eqref{3.13} obtained within the Grad method, apart from the stationarity of the system and the geometry and symmetry properties of the planar Couette flow. Now we look for hydrodynamic LTu solutions, that is,  solutions consistent with $\mathbf{q}=\text{const}$ and Eqs.\ \eqref{p}--\eqref{1}. It is easy to check that Eqs.\ \eqref{3.9}--\eqref{3.13}, together with Eq.\ \eqref{Tbal}, indeed allow for such a class of solutions. First, Eqs.\ \eqref{3.9}--\eqref{3.11} become a set of algebraic equations whose solution yields $P_{xy}/p$, $P_{xx}/p$, and $P_{yy}/p$ in terms of $\alpha$ and $a$. The reduced shear rate {$a$} is subsequently obtained as a function of $\alpha$ from Eq.\ \eqref{3.4}. Once the pressure tensor is known, Eqs.\ \eqref{3.12} and \eqref{3.13} provide $q_x/A$ and $q_y/A$ as functions of $\alpha$ for arbitrary $A$. The results can be conveniently expressed in the forms of Eqs.\ \eqref{eta}, \eqref{theta}, \eqref{lambda}, and \eqref{phi} with the following explicit expressions for the generalized transport coefficients:
\beq
\eta^*=\frac{\beta_1}{(\beta_1+\zeta^*)^2},
\label{3.14}
\eeq
\beq
\theta_x=\frac{\beta_1+d\zeta^*}{\beta_1+\zeta^*},
\label{3.15}
\eeq
\beq
\theta_y=\frac{\beta_1}{\beta_1+\zeta^*},
\label{3.15b}
\end{equation}
\begin{equation}
\label{3.16}
\lambda^*=\beta_2\frac{(d-1)(d+2)[(d+4)\theta_y-2]+d^2(d+4)(\zeta^*/\beta_2)}{(d+2)^2(d-1)\beta_2^2-
2\frac{d^2(d+4)}{d-1}a^2},
\end{equation}
\begin{equation}
\label{3.17}
\phi^*=(d+4)a\frac{d[(d+4)\theta_y-2]+(d-1)(d+2)\eta^*\beta_2}{(d+2)^2(d-1)\beta_2^2-
2\frac{d^2(d+4)}{d-1}a^2}.
\end{equation}
Here, we recall that $\eta^*=\eta/(p/\nu)$ and $\zeta^*=\zeta/\nu$. According to Eq.\ \eqref{3.4},  the dependence of the reduced shear rate $a(\alpha)$ on the coefficient of restitution $\alpha$ is
\begin{equation}
\label{3.18}
a^2=\frac{d\zeta^*}{2\beta_1}(\beta_1+\zeta^*)^2.
\end{equation}
In Eqs.\ \eqref{3.16} and \eqref{3.17} we have introduced the reduced coefficients $\lambda^*=\lambda/\lambda_\NS^0$ and $\phi^*=\phi/\lambda_\NS^0$, where
\beq
\lambda_\NS^0=\frac{d(d+2)}{2(d-1)}\frac{p}{m\nu}
\label{3.19}
\eeq
is the NS thermal conductivity in the elastic limit.
As a simple test, note that in the limit $\alpha\to 1$ {(that is, $\zeta\to 0$)} one has $a\to 0$, $\beta_i\to 1$, $\theta_i\to 1$, $\eta^*\to 1$,  $\lambda^*\to 1$, and $\phi^* \to 0$.

{}From the symmetry relation $\theta_x+\theta_y+(d-2)\theta_z=d$ and from  Eqs.\ \eqref{3.15} and \eqref{3.15b} it follows that $\theta_z=\theta_y$.
Equations (\ref{3.14})--(\ref{3.18}) extend to  {arbitrary dimensionality} $d$ our previous results for hard spheres ($d=3$) \cite{VSG10}.

The transport coefficients \eqref{3.14}--\eqref{3.17} describe the non-Newtonian properties of the granular gas in the LTu class of flows in the context of Grad's solution to the Boltzmann equation. These coefficients clearly contrast with the ones {obtained} in the NS description, where one has \cite{BDKS98,BC01}
\beq
\eta_\NS^*=\frac{1}{\beta_1+\frac{1}{2}\zeta^*},
\label{etaNS}
\eeq
\beq
\lambda_\NS^*=\frac{\beta_2-\frac{5d}{2(d-1)}\zeta^*}{\left(\beta_2-\frac{2d}{d-1}\zeta^*\right)\left(\beta_2-\frac{3d}{2(d-1)}\zeta^*\right)}.
\label{lambdaNS}
\eeq
Upon writing Eq.\ \eqref{lambdaNS} we have taken into account that the NS constitutive equation $\mathbf{q}=-\kappa_\NS \nabla T-\mu_\NS \nabla n$ becomes $\mathbf{q}=-\lambda_\NS \nabla T$, with $\lambda_\NS=\kappa_\NS-(n/T)\mu_\NS$, under the condition $\nabla p=0$.
In Eqs.\ \eqref{etaNS} and \eqref{lambdaNS}, non-Gaussian corrections to the homogeneous cooling state distribution have been neglected, in consistency with the Grad approximation \eqref{3.1}. Apart from Eqs.\ \eqref{etaNS} and \eqref{lambdaNS}, the NS description predicts $\theta_i=1$ and $\phi=0$.

\begin{figure}
  \includegraphics[width=0.95\columnwidth]{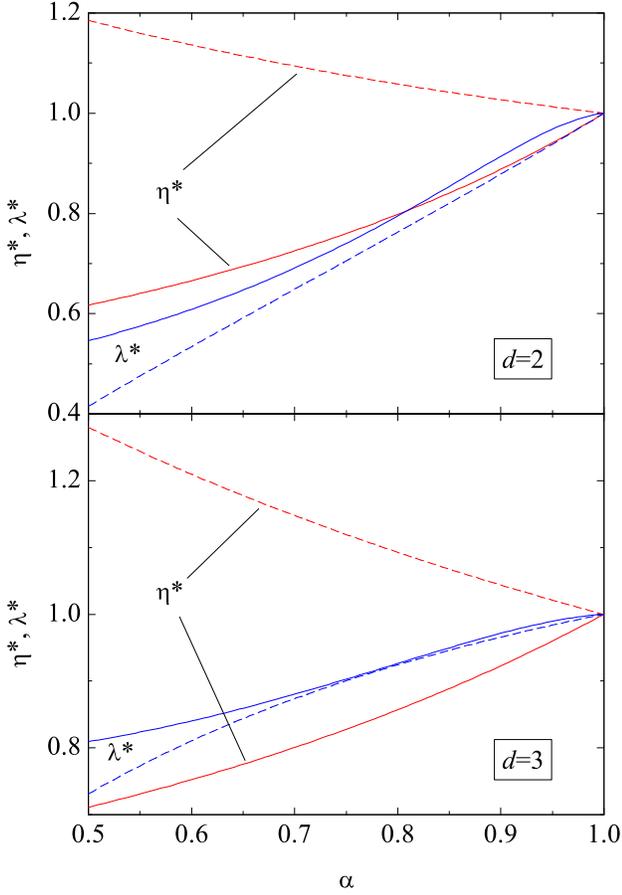}
\caption{{(Color online) Reduced shear viscosity ($\eta^*$) and thermal conductivity ($\lambda^*$) for inelastic hard disks (top panel) and hard spheres (bottom panel), as obtained from Grad's 13-moment method (solid lines) and from the NS equations (dashed lines).}}\label{fig3}
\end{figure}

Figure \ref{fig3} compares the non-Newtonian coefficients $\eta^*(\alpha)$ and $\lambda^*(\alpha)$ with their NS counterparts $\eta_\NS^*(\alpha)$ and $\lambda_\NS^*(\alpha)$ {for hard disks  ($d=2$) and hard spheres  ($d=3$). It is apparent that the LTu shear viscosity clearly differs from the NS shear viscosity. In fact, while the latter increases with increasing inelasticity, the former presents the opposite behavior \cite{SGD04}. On the other hand, both thermal conductivity coefficients are rather close to each other, especially in the case of hard spheres. It is interesting to remark that, while the NS heat-flux transport coefficients $\kappa_\NS$ and $\mu_\NS$ increase with inelasticity, the effective NS thermal conductivity $\lambda_\NS=\kappa_\NS-(n/T)\mu_\NS$ decreases. This shows the importance of the coefficient $\mu_\NS$ (absent in the elastic case) in granular flows beyond the quasi-elastic limit.}

\subsection{BGK-type kinetic model}
\label{BGK}

Now we consider the results derived for the LTu class from a BGK-type kinetic model of the Boltzmann equation \cite{BDS99}. In the geometry of the Couette flow, the steady  kinetic model becomes
\begin{equation}
\label{14} v_y\frac{\partial f}{\partial y}=-\beta(\alpha)\nu
(f-f_0)+\frac{\zeta}{2}\frac{\partial}{\partial {\bf v}}\cdot {\bf V} f,
\end{equation}
where $\nu$ is the effective collision frequency defined by Eq.\ (\ref{3.5}). The parameter $\beta(\alpha)$ is a free parameter of the model  chosen to optimize the agreement with the Boltzmann results. In terms of the variable $s(y)$ defined as $\dd s=\beta\nu(y)\dd y$, Eq.\ (\ref{14}) can be rewritten as \cite{TTMGSD01}
\beqa
\label{15}
\left(1-\frac{d}{2}\widetilde{\zeta}+{V_y}\frac{\partial}{\partial
s}-\widetilde{a}V_y\frac{\partial}{\partial V_x}\right.&-&\left.\frac{1}{2}\widetilde{\zeta}{\bf V}\cdot\frac{\partial}{\partial {\bf V}} \right)f(s,{\bf V})\nn
&&=f_0(s,{\bf V}),
\eeqa
where $\widetilde{a}\equiv a/\beta$, $\widetilde{\zeta}\equiv \zeta^*/\beta$,  and the derivative $\partial_s$ is taken at constant ${\bf V}={\bf v}-{\bf u}(s)$. Upon writing Eq.\ (\ref{15}), use has been made of Eq.\ (\ref{1}).
The hydrodynamic solution to Eq.\ (\ref{15}) is
\beqa
\label{16}
f(s,\mathbf{V})&=&\int_0^\infty \dd t\, e^{-(1-\frac{d}{2}\widetilde{\zeta})t} e^{-{\tau(t) {V_y}\partial_s}}
e^{\widetilde{a}tV_y\partial_{V_x}}\nn
&&\times f_0(s,e^{\frac{1}{2}\widetilde{\zeta}t}\mathbf{V}),
\eeqa
where
\begin{equation}
\label{17}
\tau(t)\equiv \frac{2}{\widetilde{\zeta}}\left(e^{\frac{1}{2}\widetilde{\zeta}t}-1\right).
\end{equation}
 The action of the operators $e^{-\tau
{V_y}\partial_s}$ and $e^{\widetilde{a}tV_y\partial_{V_x}}$ on an arbitrary function $g(s,{\bf V})$ is \cite{TTMGSD01}
\begin{equation}
\label{18}
e^{-{\tau \frac{V_y}{\beta}\partial_s}}g(s,{\bf V})=g(s-\tau \frac{V_y}{\beta},{\bf V}),
\end{equation}
\begin{equation}
\label{18b}
e^{\widetilde{a}tV_y\partial_{V_x}}g(s,{\bf V})=
g(s,\mathbf{V}+\widetilde{a}tV_y\widehat{\mathbf{x}}),
\end{equation}
respectively. The solution (\ref{16}) adopts the {\em normal} or hydrodynamic form since its spatial dependence only occurs through a functional dependence on the hydrodynamic fields $n(s)$, ${\bf u}(s)$, and $T(s)$ via the local equilibrium distribution $f_0$.

The objective now is two-fold. First, we want to check that the LTu profiles \eqref{p}--\eqref{1} are consistent with the solution \eqref{16}. Next, we will evaluate the fluxes and identify the generalized transport coefficients defined by Eqs.\ \eqref{eta}, \eqref{theta}, \eqref{lambda}, and \eqref{phi}. In order to accomplish this two-fold objective, it is convenient to define  the {general} velocity moments
\beq
M_{k_1,k_2,k_3}(s)=\int \dd\mathbf{V}\, V_x^{k_1}V_y^{k_2}V_z^{k_3}f(s,\mathbf{V}).
\label{3.20}
\eeq
Insertion of Eq.\ \eqref{16} yields
\begin{widetext}
\beqa
M_{k_1,k_2,k_3}(s)&=&\int_0^\infty  \dd t\int \dd\mathbf{V} e^{-(1-\frac{d}{2}\widetilde{\zeta})t}\left(V_x-\widetilde{a}t V_y\right)^{k_1} V_y^{k_2}V_z^{k_3} e^{-{\tau(t) {V_y}\partial_s}}
f_0(s,e^{\frac{1}{2}\widetilde{\zeta}t}\mathbf{V})\nn
&=&\int_0^\infty  \dd t\int \dd\mathbf{V} e^{-(1+\frac{k}{2}\widetilde{\zeta})t}\left(V_x-\widetilde{a}t V_y\right)^{k_1} V_y^{k_2}V_z^{k_3} e^{-{\tau_1(t) {V_y}\partial_s}}
f_0(s,\mathbf{V}),
\label{3.21}
\eeqa
%\end{widetext}
where $k\equiv k_1+k_2+k_3$ and $\tau_1(t)\equiv \tau(t)e^{-\frac{1}{2}\widetilde{\zeta}t}={2}\left(1-e^{-\frac{1}{2}\widetilde{\zeta}t}\right)/{\widetilde{\zeta}}$.
It is now convenient to expand the operator $e^{-{\tau_1(t) {V_y}\partial_s}}$, so that Eq.\ \eqref{3.21} becomes
\beqa
M_{k_1,k_2,k_3}(s)&=& \sum_{\ell=0}^{k_1} \binom{k_1}{\ell}\sum_{h=0}^\infty \frac{1}{h!}
\int_0^\infty \dd t\,e^{-(1+\frac{k}{2}\widetilde{\zeta})t} \left[-\tau_1(t)\right]^h \left(-\widetilde{a}t\right)^{k_1-\ell}  \partial_s^h\int \dd\mathbf{V}  V_x^{\ell}  V_y^{k_1+k_2-\ell+h}V_z^{k_3}
f_0(s,\mathbf{V})\nn
&=& \sum_{\ell=0}^{k_1} \binom{k_1}{\ell}\sum_{h=0}^\infty \frac{ C_{\ell}C_{k+h-\ell-k_3}C_{k_3}}{h!}A_{k,h,k_1-\ell}\partial_s^{h}\left[ n(s) \left(\frac{2T(s)}{m}\right)^{(k+h)/2}\right],
\label{3.23}
\eeqa
\end{widetext}
where

\beq
C_{\ell}=\begin{cases}
\pi^{-1/2} \Gamma\left(\frac{\ell+1}{2}\right),&\ell=\text{even},\\
0,&\ell=\text{odd},
\end{cases}
\label{3.24}
\eeq
and
\beq
A_{k,h,k_1}\equiv\int_0^\infty \dd t\,e^{-(1+\frac{k}{2}\widetilde{\zeta})t} \left[-\tau_1(t)\right]^h \left(-\widetilde{a}t\right)^{k_1}.
\label{3.25}
\eeq
In particular, $A_{0,0,0}=1$,
\beq
A_{2,0,0}=\frac{1}{1+\widetilde{\zeta}}, \quad A_{2,0,1}=-\frac{\widetilde{a}}{(1+\widetilde{\zeta})^{2}},
\label{3.26}
\eeq
\beq
A_{2,0,2}=\frac{2\widetilde{a}^2}{(1+\widetilde{\zeta})^{3}},
\eeq
\beq
A_{3,1,0}=-\frac{2}{(1+2\widetilde{\zeta})(2+3\widetilde{\zeta})},
\label{3.27}
\eeq
\beq
A_{3,1,1}=\frac{2\widetilde{a}(4+7\widetilde{\zeta})}{(1+2\widetilde{\zeta})^2(2+3\widetilde{\zeta})^2},
\label{3.27b}
\eeq
\beq
A_{3,1,2}=-\frac{4\widetilde{a}^2(12+42\widetilde{\zeta}+37\widetilde{\zeta}^2)}{(1+2\widetilde{\zeta})^3(2+3\widetilde{\zeta})^3},
\label{3.28}
\eeq
\beq
A_{3,1,3}=\frac{12\widetilde{a}^3(4+7\widetilde{\zeta})(8+28\widetilde{\zeta}+25\widetilde{\zeta}^2)}{(1+2\widetilde{\zeta})^4(2+3\widetilde{\zeta})^4}.
\label{3.29}
\eeq
Note that because of the parity properties of the coefficients $C_{\ell}$, only the terms with $\ell=\text{even}$ and $h+k=\text{even}$ contribute to the summations in Eq.\ \eqref{3.23}. Moreover, the moments $M_{k_1,k_2,k_3}$ with $k_3=\text{odd}$ vanish.

So far, no specific  spatial dependence of density and temperature has been assumed. Only the linear $s$ dependence of the flow velocity has been used. Now, we assume that $n(s)T(s)=\text{const}$ and $\partial_s T(s)=\text{const}$, in agreement with Eqs.\ \eqref{p} and \eqref{1bis}, respectively. These assumptions imply that $\partial_s^h [T(s)]^{(k+h-2)/2}=0$ if $h>(k+h-2)/2$. Therefore, the summation $\sum_{h=0}^\infty$ can be replaced by $\sum_{h=0}^{\max(0,k-2)}$ and Eq.\ \eqref{3.23} reduces to
\begin{widetext}
\beq
M_{k_1,k_2,k_3}(s)=n(s)\left[\frac{2 T(s)}{m}\right]^{k/2}\sum_{\ell=0}^{k_1} \binom{k_1}{\ell}\sum_{h=0}^{\max(0,k-2)} \frac{ C_{\ell}C_{k+h-\ell-k_3}C_{k_3}}{h!}A_{k,h,k_1-\ell}
\frac{\left(\frac{k+h}{2}-1\right)!}{\left(\frac{k-h}{2}-1\right)!}\left(\sqrt{\frac{2}{mT}}\partial_s T\right)^h.
\label{3.30}
\eeq
\end{widetext}
It is straightforward to check that $M_{0,0,0}(s)=n(s)$ and $M_{1,0,0}=M_{0,1,0}=M_{0,0,1}=0$. This proves the consistency of the assumed density and velocity profiles in the LTu flow. The consistency condition for the  temperature is $M_{2,0,0}+M_{0,2,0}+(d-2)M_{0,0,2}=dp/m$. It can be checked that this condition is satisfied provided that the reduced shear rate $\widetilde{a}$ is related to the coefficient of restitution {by}
\beq
\widetilde{a}^2=\frac{d}{2}\widetilde{\zeta}(1+\widetilde{\zeta})^2.
\label{3.31}
\eeq
This result is fully equivalent to Grad's prediction \eqref{3.18}, except that $\beta_1$ is replaced by $\beta$.

Once we have proven that the BGK-type kinetic equation \eqref{14} admits an exact solution characterized by the LTu hydrodynamic fields, we can obtain all the velocity moments from Eq.\ \eqref{3.30}. The relevant elements of the pressure tensor are $P_{xx}=mM_{2,0,0}$, $P_{yy}=mM_{0,2,0}$, and $P_{xy}=mM_{1,1,0}$. {}From them one can easily identify the dimensionless coefficients defined by Eqs.\ \eqref{eta} and \eqref{theta}. The resulting expressions coincide with Grad's results \eqref{3.14}--\eqref{3.15b}, again with the replacement $\beta_1\to\beta$.

The two non-zero components of the heat flux are $q_x=(m/2)\left[M_{3,0,0}+M_{1,2,0}+(d-2)M_{1,0,2}\right]$ and $q_y=(m/2)\left[M_{1,2,0}+M_{0,3,0}+(d-2)M_{0,1,2}\right]$. As expected, they are proportional to the temperature gradient and this allows one to identify the generalized thermal conductivities defined in Eqs.\ \eqref{lambda} and \eqref{phi}.
After some algebra, one gets
\begin{equation}
\label{21b}
\lambda^*=\frac{2/\beta}{(1+2\widetilde{\zeta})(2+3\widetilde{\zeta})}\left[1+\frac{6\widetilde{a}^2}{d+2}\frac{12+42\widetilde{\zeta}+
37\widetilde{\zeta}^2}{(1+2\widetilde{\zeta})^2(2+3\widetilde{\zeta})^2}\right],
\end{equation}
\beqa
\label{22}
\phi^*&=&\frac{2\widetilde{a}}{d+2}\frac{4+7\widetilde{\zeta}}{(1+2\widetilde{\zeta})^2(2+3\widetilde{\zeta})^2}\nn
&&\times
\left[d+4+
18\widetilde{a}^2\frac{8+28\widetilde{\zeta}+
25\widetilde{\zeta}^2}{(1+2\widetilde{\zeta})^2(2+3\widetilde{\zeta})^2}\right],
\eeqa
where we have taken into account that in the BGK model the NS thermal conductivity in the elastic case is not given by Eq.\ \eqref{3.19} but by $\lambda_\NS^0=\frac{d+2}{2}p/m\nu$.
Comparison with Eqs.\ \eqref{3.16} and \eqref{3.17} shows that the transport coefficients $\lambda$ and $\phi$ predicted by the BGK model  are  different from those obtained from Grad's method, regardless of the choice of the free parameter $\beta$.

So far, $\beta$ has remained free. Henceforth, by following arguments presented in Refs.\ \cite{SA05} and \cite{VGS07}, we will take, for simplicity, $\beta=(1+\alpha)/2$.

\section{Simulation methods}
\label{sec4}
{As said in the Introduction, in order to assess the reliability of the previously described  theoretical results and the existence of the LTu class, we} have performed DSMC simulations of the {Boltzmann equation} and MD simulations  for a granular gas of hard spheres {($d=3$)} {\cite{LVU09}. In the MD simulations  the global solid {volume} fraction has been taken equal to $7\times 10^{-3}$ in order to remain in the dilute regime and compare with the Boltzmann results obtained either from DSMC simulations or from the theoretical approaches.} The gas is enclosed between two {plates moving with velocities $U_\pm$ and maintained} at temperatures $T_\pm$, where {the subscripts} $+$ and $-$  denote upper and lower wall, respectively (see Fig. \ref{fig1}).

{In our simulations we have considered $N=2\times 10^5$ particles (DSMC) and $N\sim 10^4$--$10^5$ particles (MD). When} a particle collides with a wall its velocity is updated following the rule $\mathbf {v}\rightarrow\mathbf{v}'+U_\pm{\widehat{\mathbf{x}}}$. The first {contribution} ($\mathbf{v}'$) of the new particle velocity is due to thermal boundary condition, {while} the second {contribution} ($U_\pm{\widehat{\mathbf{x}}}$) is due to wall {motion}. The horizontal components of $\mathbf{v}'$ are randomly drafted from a Maxwell distribution (at a temperature $T_\pm$) whereas the normal component {$v_y'$}, due to collision with a wall, is sampled from a Rayleigh probability distribution: {$P(|v_y'|)=(m|v_y'|/T_{\pm})e^{-m{v_y'}^2/2T_\pm}$}.

{In the traditional DSMC method \cite{B94}, which we use here,  the system is split into cells whose characteristic length is much smaller than the mean free path $\ell$ (that is, macroscopic properties do not vary significantly along a cell). Here we define the (local) mean free path for hard spheres as $\ell=\sqrt{T/m}\nu^{-1}$, where the (local) effective collision frequency $\nu$ is defined in Eq.\ \eqref{3.5}. Furthermore, the time step  needs to be much smaller than the microscopic characteristic time (inverse of the collision frequency $\nu$). The DSMC method}
consists of two steps. One is the free streaming, where the particles move in straight lines without inter-particle collisions. The boundary conditions are applied in this step. The other {one} is the collision step, in which possible particle pairs are randomly selected from the same cell and collision is accepted with a probability $\Theta(\mathbf{v}_{ij}\cdot\widehat{\bm{\sigma}}_{ij})\omega_{ij}/\omega_{\mathrm{max}}$, where $\mathbf{v}_{ij}=\mathbf{v}_i-\mathbf{v}_j$ is the relative velocity between particles $i$ and $j$, {$\widehat{\bm{\sigma}}_{ij}=(\mathbf{r}_i-\mathbf{r}_j)/|\mathbf{r}_i-\mathbf{r}_j|$}, $\omega_{ij}=(4\pi\sigma^2n)|\mathbf{v}_{ij}\cdot\widehat{\bm{\sigma}}_{ij}|$, {and} $\omega_{\mathrm{max}}$ is an upper bound of the probability of particle collision per unit time.

{Given the geometry of the problem,} the DSMC cells need not be {three-dimensional} since only the vertical coordinate $y$ is recorded. This is possible {because} collisions are sampled independently of the particle position {within the same layer}, and only relative approach velocities $\mathbf{v}_{ij}\cdot\widehat{\bm{\sigma}}_{ij}$ are needed in the simulation (unit vectors $\widehat{\bm{\sigma}}_{ij}$ are randomly generated).
In our DSMC simulations we have taken a time step and a layer width  given by $\delta t=3\times 10^{-3}\bar\nu^{-1}$ and $\delta y =2\times 10^{-2}\bar\ell$, respectively, {where} (as said in Sec.\ \ref{Rey})
$\bar\ell=\sqrt{T_-/m}\bar\nu^{-1}$,  $\bar\nu$ being given by Eq.\ \eqref{3.5} with $T\to T_-$ and $n\to\bar{n}$.

In contrast to the DSMC case, a three-dimensional box is required in the MD simulations. We have taken $h\times h\times h$ cubes with periodic boundary conditions along the  directions ($x$ and $z$) parallel to the walls.

{In the simulation results presented in Sec.\ \ref{sec5} dimensionless quantities are used}. We choose as units {of mass, length, and time $m$, $\ell(-h/2)$, and $\nu^{-1}(-h/2)$, respectively, once the steady state has been reached}. As said before, we take the condition $T_-\leq T_+$. Thus, and with our choice of units, the reduced quantity {$A/\sqrt{mT(-h/2)}$ [cf.\ Eq.\ \eqref{1bis}]} will represent the \emph{maximum} value across the system of the {local} thermal Knudsen number $\epsilon$  {[cf.\ Eq.\ \eqref{epsilon}]}. In other words, {in} our work the slope of the thermal Knudsen number $\epsilon(y)$ is always positive \cite{VU09}.

The separation between the plates has typically been set $h\approx 5$--$20$  and we have considered a wall temperature difference in the range $\Delta T\equiv T_+/T_--1=0$--$20$.
Since, as will be seen below [cf.\ Fig.\ \ref{fig9}(a)], the values of the reduced shear rate are smaller than $1$ for $\alpha\geq 0.5$, the above values of $h$ and $\Delta T$ imply that the Reynolds number [cf.\ Eq.\ \eqref{R3}] is always smaller than about $400$. For this range of $\text{Re}$ the flow is expected to remain laminar and this is confirmed by our simulations.

{We store instantaneous values of the relevant hydrodynamic quantities iteratively at runtime, for further processing after the simulation run.}

With respect to the processing of the steady state hydrodynamic properties, we perform two types of averages: one in space and the second one in time. The first one is performed over a number of contiguous cells, forming a statistical spatial bin, whose size must not be larger than the typical scale over which hydrodynamic fields vary \cite{VGS08}.  Since this scale depends on the applied gradients (wall temperature difference and applied shear for a system with a given height $h$), this statistical bin size needs to be adjusted for each simulation. We have observed that a bin adjustment of $\Delta y \approx 0.1\mathrm{Kn}^{-1}\bar\ell$ (where the Knudsen number is $\text{Kn}=a$) is enough for preserving all properties of hydrodynamic profiles \cite{VU09,VGS08}. The other averaging is performed, in each spatial bin, over values at different times of the same steady states. This double averaging is very convenient since it allows us to obtain very smooth hydrodynamic steady profiles, even if the system is not large. This is {especially} useful in the case of DSMC, where thermal fluctuations may result in too noisy profiles for small systems \cite{B94,HGBH03}.

\begin{figure}
  \includegraphics[width=0.95\columnwidth]{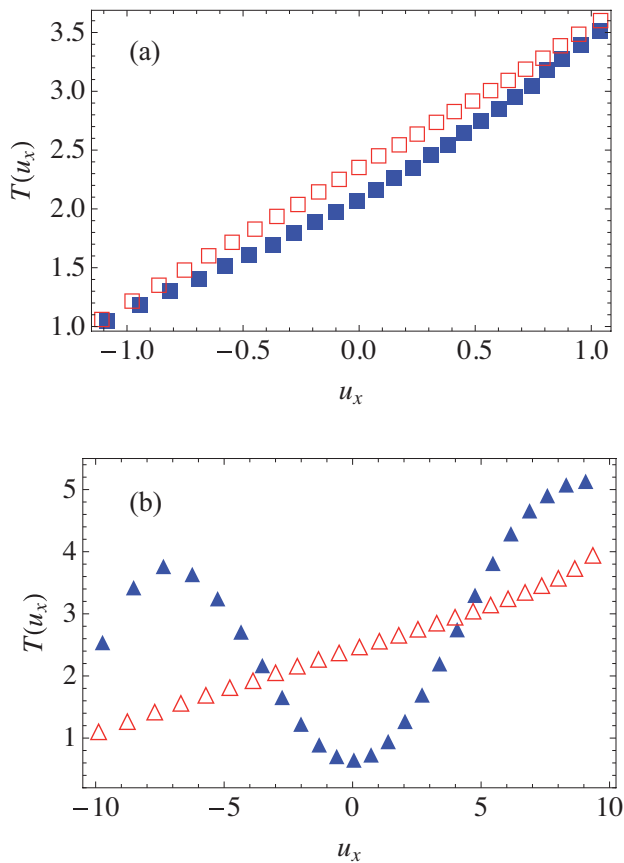}
\caption{(Color online) {Temperature  vs flow velocity, $T(u_x)$, as obtained from DSMC simulations at {$t=45\bar\nu^{-1}$} (solid symbols, transient state) and {$t>800\bar\nu^{-1}$} (open symbols, steady state). In these graphs $\Delta T=4$ and (a)  $\alpha=0.99$ and (b) $\alpha=0.4$.} }\label{fig4}
\end{figure}

\begin{figure}[h]
  \includegraphics[width=0.95\columnwidth]{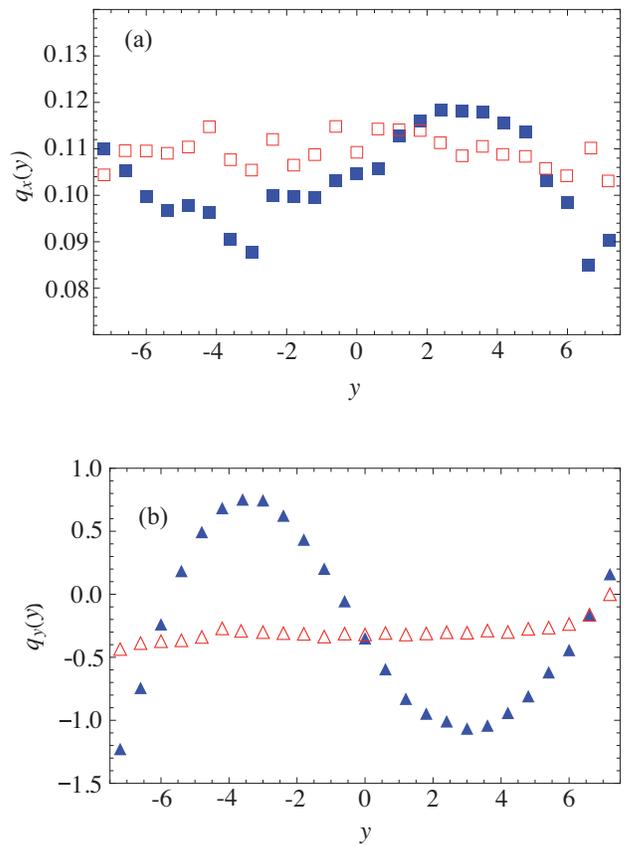}
\caption{(Color online) {Heat flux  profiles (a) $q_x(y)$ and (b) $q_y(y)$ as obtained from DSMC simulations at {$t=45\bar\nu^{-1}$} (solid symbols, transient state) and {$t>800\bar\nu^{-1}$} (open symbols, steady state). In these graphs $\Delta T=4$ and (a)  $\alpha=0.99$ and (b) $\alpha=0.4$.}}\label{fig5}
\end{figure}

\section{Results}
\label{sec5}

\subsection{Transient regime}

We analyze here the transition to steady LTu states from DSMC and MD simulation data, starting from an initial equilibrium distribution at $T=T_-$. {We have found that in general the duration of this transition to the steady state becomes  substantially longer as inelasticity increases.}

\begin{figure}
  \includegraphics[width=0.95\columnwidth]{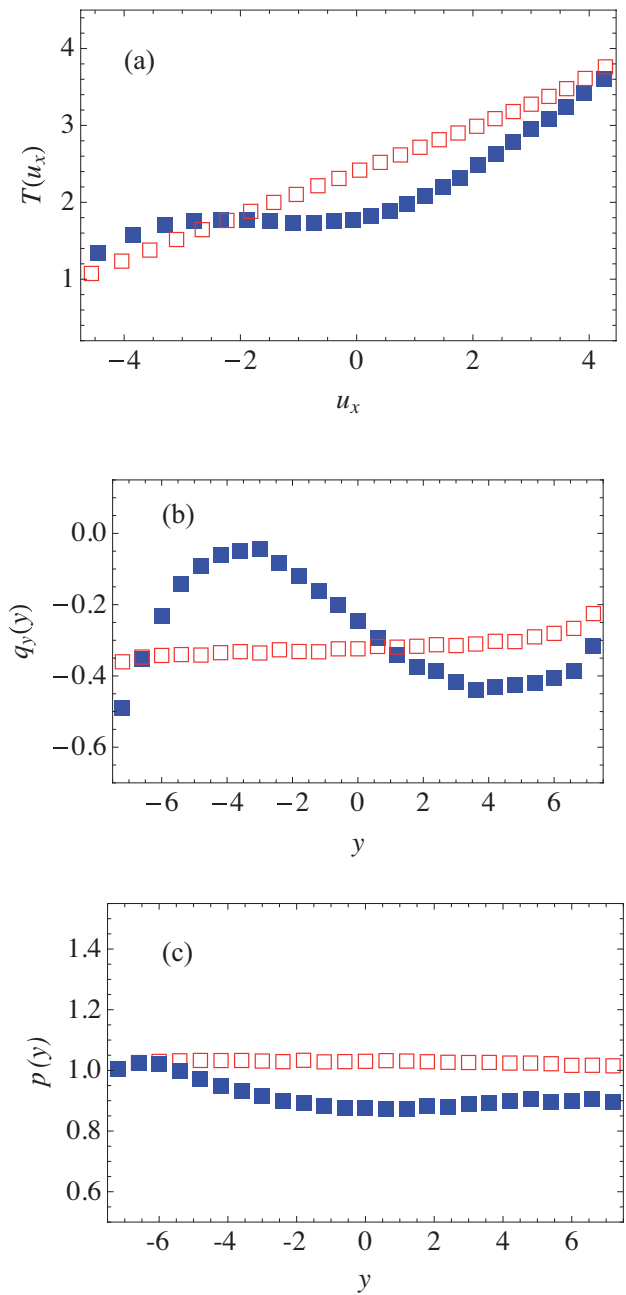}
\caption{(Color online) {Plots of  (a) $T(u_x)$, {(b) $q_y(y)$, and (c) $p(y)$} as obtained from MD simulations at {$t=60\bar\nu^{-1}$} (solid symbols, transient state) and {$t>800\bar\nu^{-1}$} (open symbols, stationary state). In these graphs $\Delta T=4$ and  $\alpha=0.85$.}} \label{fig6}
\end{figure}

Figures \ref{fig4} and \ref{fig5} show $T(u_x)$ and $q_{x,y}(y)$ profiles, respectively,  from DSMC data for transient states at {$t=45\bar\nu^{-1}$} and steady states {($t>800\bar\nu^{-1}$)} for $\alpha=0.99$  and $0.4$. {In these cases $h\simeq 16$, as indicated by the horizontal axis of Fig.\ \ref{fig5}.} {It is apparent that, at a given common time, the  deviations from the steady LTu  profiles are  weaker for  $\alpha=0.99$ (quasi-elastic gas) than for  $\alpha=0.4$ (strongly inelastic gas)}. In any case, we have seen that over the range of $\alpha$ at which we perform the simulations ($\alpha=0.3$--$1.0$), time values of about {$t=250\bar\nu^{-1}$} always yield fully developed steady LTu flows. This happens also for MD simulations, as shown in Fig.\ \ref{fig6}, where we can see results for {temperature, heat flux, and pressure} for a granular gas with $\alpha=0.85$. The degree of approach to the steady state is perhaps a little {slower} but, in any case,  {we have observed that the system has already reached the steady state at {$t=250\bar\nu^{-1}$.}}

{Figures \ref{fig4}--\ref{fig6} show that both DSMC and MD confirm the existence, in the steady state,  of  Couette flows with practically linear $T(u_x)$ profile, uniform heat flux, and uniform pressure.}

\subsection{Identification of LTu flows}

In order to identify the LTu flows, we {have proceeded} analogously to a previous work {\cite{VSG10}}. For each simulation series, we fix $\Delta T$ and the applied shear $(U_{+}-U_{-})/h$. Once the steady state is reached, we monitor the parametric plot of temperature versus flow field, $T(u_x)$, looking for the typical {linear} profiles of the LTu steady states {in the \emph{bulk} region, that is, outside the boundary layers}. More specifically, since we observed that $T(u_x)$ never shows inflection points in the bulk (in accordance with theory \cite{VU09}), we check the sign {of the} $T(u_x)$ profile curvature. If {the sign} is positive, that means that cooling still overcomes viscous heating. {Thus}, we need still {increase  the applied shea}r for the next simulation (while keeping  constant $\Delta T$), in the search for a $T(u_x)$ profile with zero curvature. The process is repeated iteratively until we observe a change of sign in the curvature of $T$. Then, we {look for} the LTu state between the consecutive values for which the change of sign in the curvature is observed, by taking smaller changes of applied shear  and looking at both $T(u_x)$ and $q_{x,y}$. We take as the {final} LTu flow the simulation which best approaches the conditions of both linear $T(u_x)$ and constant $q_{x,y}$ ($q_z$ is always zero in our geometry). Put in other words, we find the LTu flows by crossing vertically (in the shearing axis direction) the surface in Fig.\ \ref{fig2}, {until getting the right value of the reduced shear rate ($a_\ltu$)}.

\begin{figure}
  \includegraphics[width=0.95\columnwidth]{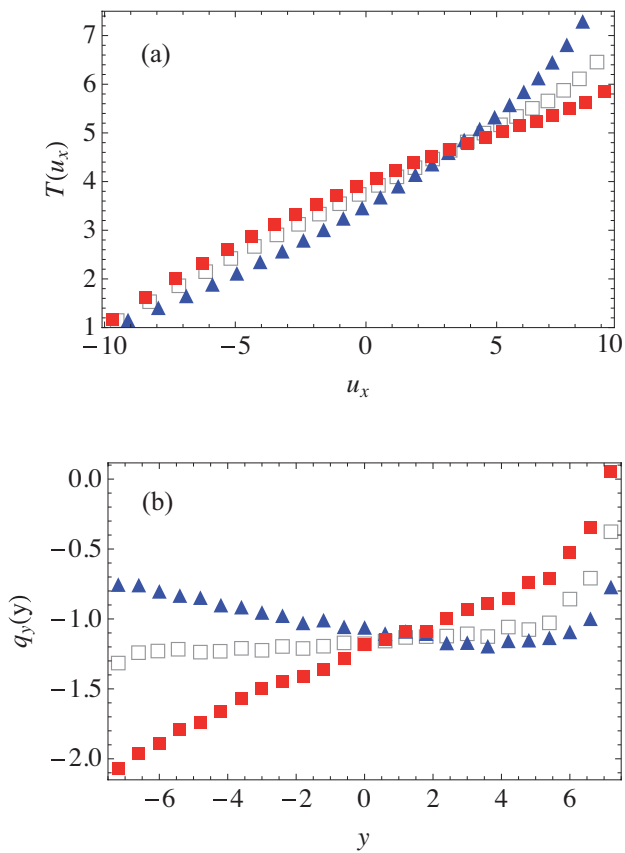}
\caption{(Color online) Transition to LTu profiles for MD series with
  varying wall shearing {at $\alpha=0.6$}. Solid symbols correspond to non-LTu states: ($\blacktriangle$) for $a=0.87a_{\ltu}$  and ($\blacksquare$) for $a=1.25a_{\ltu}$). Open squares ($\square$) stand for the LTu stationary profile.}\label{fig7}
\end{figure}

The degree of approach to the properties of theoretical LTu states that we obtained in the simulations is rather good. For illustration on this, we show MD simulation results in Fig.\ \ref{fig7}, where one can see how the transition between states above and below the surface in Fig.\  \ref{fig2} occurs. {Figure \ref{fig7}(a) shows the results for $T(u_x)$ profiles, whereas Fig.\ \ref{fig7}(b) shows} the corresponding results for $q_y(y)$ profiles. We have found that heat flux profiles are more sensitive to a departure from the LTu surface and for this reason we usually proceed as described above:  we first search for an almost linear $T(u_x)$ profile and then we fine tune the LTu state by searching the flattest heat flux profiles for a shear rate around the first selected value. Compared to results from DSMC simulations (see Fig.\ 2(a) in Ref.\ \cite{VSG10}) we see that boundary layer effects on heat flux profiles are stronger in MD simulations. {Also, this effect is more noticeable next to the higher temperature wall}. It is also to be noticed that the {sign of $\partial_y q_y(y)$ in the bulk domain changes from positive for $a>a_{\ltu}$ to negative for $a<a_{\ltu}$.} This agrees with the interpretation that viscous heating carries kinetic energy toward the hotter wall whereas inelastic cooling tends to remove it from there \cite{VU09}. Once this effect of inelastic cooling is sufficiently compensated by viscous heating, we can see the traditional trend of heat flux profiles for elastic gases between two walls at different temperatures (that is, heat flux is directed toward the colder wall \cite{B94}).

\begin{figure}
  \includegraphics[width=0.95\columnwidth]{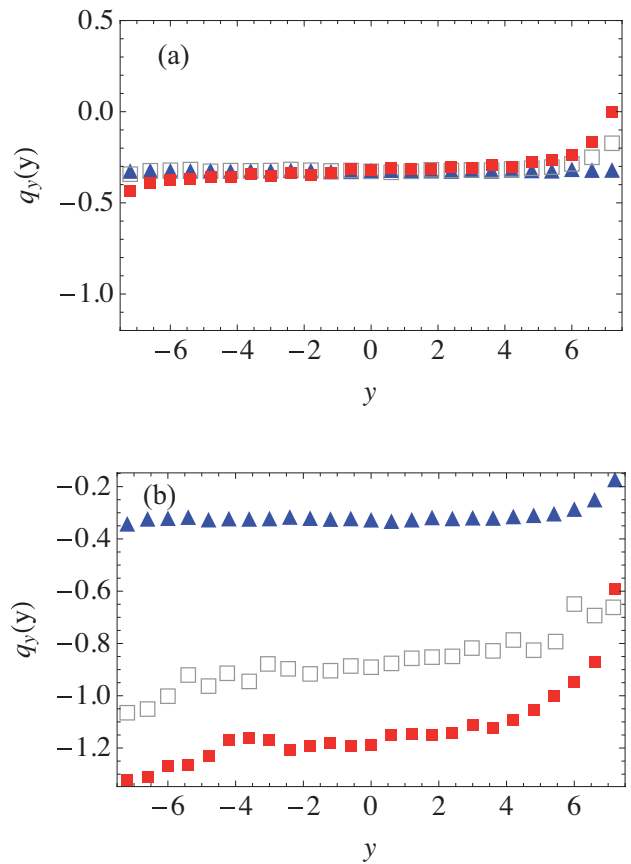}
\caption{(Color online) {Heat flux profiles $q_y(y)$ from DSMC data. In panel (a)
 $\Delta T=5$ and $\alpha=0.4$ ($\square$),  $\alpha=0.7$ ($\blacksquare$), and
   $\alpha=0.99$ ($\blacktriangle$). In panel (b) $\alpha=0.7$ and $\Delta T=5$ ($\blacktriangle$),  $\Delta T=10$  ($\square$), and  $\Delta T=15$ ($\blacksquare$).}}\label{fig8}
\end{figure}

{In Fig.\ \ref{fig8} we show LTu heat flux profiles from DSMC data. It is observed that, at a given wall temperature difference $\Delta T$, the impact of $\alpha$ on $q_y$ is rather weak. On the other hand, at a given value of $\alpha$, the magnitude of $q_y$ is approximately proportional to $\Delta T$. Although not shown, we have also found that the influence of $\alpha$ on $q_x$ is much stronger than in the case of $q_y$.}

\subsection{Generalized transport coefficients}

In a recent work {\cite{VSG10}}, we introduced the method of measurement of the {generalized} transport coefficients of the LTu class {defined by Eqs.\ \eqref{eta}, \eqref{theta}, \eqref{lambda}, and \eqref{phi}. We have confirmed by simulations that the values of these reduced  coefficients  only depend} on the value of the coefficient of normal restitution $\alpha$.

\begin{figure}
  \includegraphics[width=0.95\columnwidth]{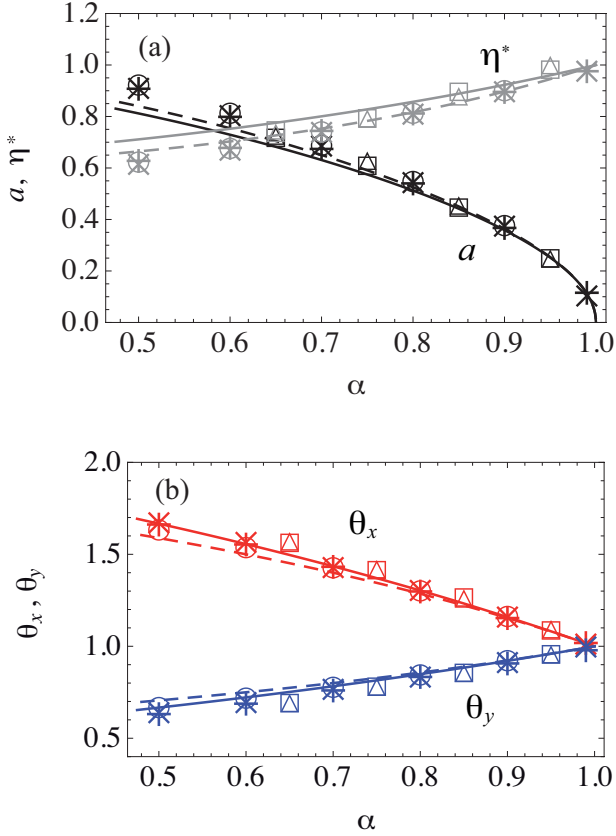}
\caption{(Color online)  {Plot of (a) $a(\alpha)$ and $\eta^*(\alpha)$ and (b) $\theta_x(\alpha)$ and $\theta_y(\alpha)$ as obtained from DSMC simulations ($h=15$) with  $\Delta T=0$ ($\bigcirc$) (USF data from Ref.\ \protect\cite{AS05}), $\Delta T=2$  ($\times$), and $\Delta  T=10$   ($+$), and from MD simulations ($h=7$) with $\Delta T=2$  ($\triangle$) and $\Delta  T=5$   ($\square$).
The solid and dashed lines correspond to Grad's method and BGK model, respectively.}}\label{fig9}
\end{figure}

{Figure \ref{fig9} presents the simulation data for the reduced shear rate $a$, the reduced shear viscosity $\eta^*$, and the reduced directional temperatures $\theta_i$ as functions of the coefficient of normal restitution $\alpha$. The figure also includes the theoretical predictions obtained from Grad's method [cf.\ Eqs.\ \eqref{3.14}--\eqref{3.15b} and \eqref{3.18} with $\beta_1$ given by Eq.\ \eqref{3.6}] and from the BGK-like kinetic model [cf.\ Eqs.\ Eqs.\ \eqref{3.14}--\eqref{3.15b} and \eqref{3.18} with the replacement $\beta_1\to (1+\alpha)/2$].
It can be observed a consistent agreement between DSMC and MD data. Moreover, the theoretical results compare quite well with computer simulations, the BGK results slightly improving the results obtained from Grad's approximation.}

\begin{figure}
  \includegraphics[width=0.95\columnwidth]{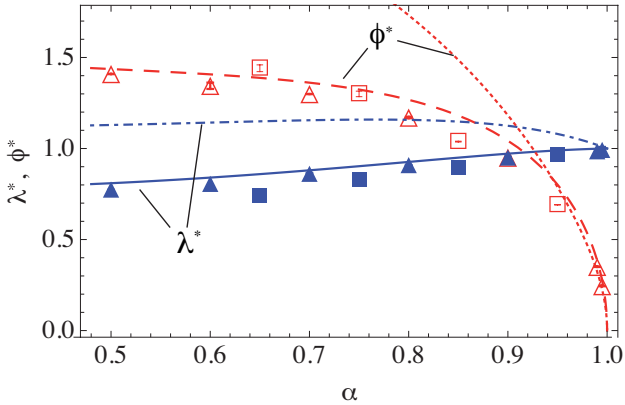}
\caption{(Color online) {Plot of $\lambda^*(\alpha)$ ($\blacktriangle$, $\blacksquare$) and $\phi(\alpha)$ ($\triangle$, $\square$) as obtained from DSMC simulations (triangles)  and from MD simulations (squares).
The solid ($\lambda^*$) and dashed ($\phi^*$) lines correspond to Grad's method, while the dotted-dashed ($\lambda^*$)   and dotted lines ($\phi^*$) stand for the BGK model.}}\label{fig10}
\end{figure}

{Regarding the transport coefficients characterizing the heat flux, Fig.\ \ref{fig10} compares computer simulation results (DSMC and MD) with Grad's [cf.\ Eqs.\ \eqref{3.16} and \eqref{3.17}  with $\beta_2$ given by Eq.\ \eqref{3.8}] and BGK [cf.\ Eqs.\ \eqref{21b} and \eqref{22}] theoretical predictions.
It is apparent that the generalized thermal conductivity $\lambda^*$ exhibits a  weak dependence on $\alpha$, in agreement with Fig.\ \ref{fig8}(a). On the other hand, the cross coefficient $\phi^*$, which vanishes in the elastic limit, starts growing rapidly with increasing inelasticity, and then presents a much more moderate dependence on $\alpha$ for large inelasticities.
In particular, $\phi^*$ becomes larger than $\lambda^*$ for $\alpha\lesssim 0.9$, what represents a strong non-Newtonian effect.
Interestingly, these features are very well captured by the simple Grad approximation, while the BGK approach only agrees at a qualitative level. The contrast between the good performance of the BGK predictions for the rheological properties seen in Fig.\ \ref{fig9} and the quantitative disagreement found in Fig.\ \ref{fig10} is in part due to the fact that the BGK model only possesses a free parameter ($\beta$) to make contact with the Boltzmann equation.}

\section{{Concluding remarks}}
\label{sec6}

We have {presented} in this work {an extensive study of a} class of granular flows recently reported \cite{VSG10}. We {refer to} this class of flows  as 'LTu,'  due to the linearity of $T(u_x)$ profiles. Our study {has been} both theoretical and computational. In the theory part, we have presented results from two different approaches: Grad's moment method and a BGK-type kinetic model used previously in other granular flow problems and now applied specifically to the LTu flows. In the computational part, we {have presented} results also from two different methods: {the DSMC method} of the Boltzmann equation of the inelastic gas and {MD simulations of a dilute gas}.

{The objective of the paper has been two-fold. First, we have confirmed by computer simulations the existence of LTu flows in the bulk domain under strongly inelastic conditions. At a given wall temperature difference and by a careful fine-tuning of the  shear rate applied by the walls, it is possible to reach steady states with a uniform heat flux and a linear parametric plot of $T$ vs $u_x$. Second, we have assessed the theoretical predictions derived from two different approaches (Grad's moment method and BGK-type kinetic model) for the generalized non-Newtonian transport coefficients.}

The agreement for the reduced shear rate, rheological properties, and transport coefficients between the DSMC and MD simulation methods is very good, as  shown in Figs.\ \ref{fig9} {and \ref{fig10}}. Also, the evolution to stationary states and other properties of the hydrodynamics of the LTu class are found to be remarkably similar for both DSMC and MD. {Regarding the reliability of both theoretical solutions, we have observed that {they} are excellent for the rheological properties [cf.\ Fig.\ \ref{fig9}].  On the other hand, in the case of the heat flux coefficients, the quantitative agreement with simulation is only good for Grad's moment method.}
This good performance of Grad's method has been also observed in the case of granular binary mixtures under simple shear flow \cite{MG02,L04}. {Nevertheless}, the good behavior of  Grad's 13-moment method does not extend to cases where the heat flux is not  uniform, as  happens in the Couette flow for ordinary gases \cite{GS03}.

As {it} is customary in fluid mechanics, {the importance of describing entire classes of flows with clearly identifiable hydrodynamic properties (rather than describing specific properties of a given flow in a case-by-case basis) cannot be overemphasized}. In this sense, we have shown here that the LTu flows are characterized by a set of interesting properties that can be useful as a reference point for experimental studies on granular flow at low density. More interestingly, we show that all flows of the new class share, for the same $\alpha$, the same Knudsen number associated with transport of momentum.

To summarize, we have described in detail the properties of a new class of flows, finding excellent agreement between simulation and theory. The results show that this class of flows encompasses at the same time flows of elastic and inelastic gases, what gives solid support to the validity of a hydrodynamic description of granular dynamics, at least in this case and for the type of geometry studied in this work.

We expect in the future to extend these results {to other related systems, such as mixtures, inelastic rough spheres, or driven systems. Also, we plan to carry out} further studies on the hydrodynamics of this type of flows (instabilities, pattern formation, etc.).

\acknowledgments
This research has been supported by the Ministerio de  Ciencia e Innovaci\'on (Spain) through Grant No. FIS2010-16587 (partially financed by FEDER funds).

\bibliography{long_LTu}

\begin{thebibliography}{52}
\expandafter\ifx\csname natexlab\endcsname\relax\def\natexlab#1{#1}\fi
\expandafter\ifx\csname bibnamefont\endcsname\relax
  \def\bibnamefont#1{#1}\fi
\expandafter\ifx\csname bibfnamefont\endcsname\relax
  \def\bibfnamefont#1{#1}\fi
\expandafter\ifx\csname citenamefont\endcsname\relax
  \def\citenamefont#1{#1}\fi
\expandafter\ifx\csname url\endcsname\relax
  \def\url#1{\texttt{#1}}\fi
\expandafter\ifx\csname urlprefix\endcsname\relax\def\urlprefix{URL }\fi
\providecommand{\bibinfo}[2]{#2}
\providecommand{\eprint}[2][]{\url{#2}}

\bibitem[{\citenamefont{Boltzmann}(1995)}]{B95}
\bibinfo{author}{\bibfnamefont{L.}~\bibnamefont{Boltzmann}},
  \emph{\bibinfo{title}{Lectures on Gas Theory}} (\bibinfo{publisher}{Dover},
  \bibinfo{year}{1995}).

\bibitem[{\citenamefont{Maxwell}(1867)}]{M67}
\bibinfo{author}{\bibfnamefont{J.~C.} \bibnamefont{Maxwell}},
  \bibinfo{journal}{Phil. Trans. Roy. Soc.} \textbf{\bibinfo{volume}{157}},
  \bibinfo{pages}{49} (\bibinfo{year}{1867}).

\bibitem[{\citenamefont{Hilbert}(1912)}]{Hilbert}
\bibinfo{author}{\bibfnamefont{D.}~\bibnamefont{Hilbert}},
  \bibinfo{journal}{Math. Ann.} \textbf{\bibinfo{volume}{72}},
  \bibinfo{pages}{562} (\bibinfo{year}{1912}).

\bibitem[{\citenamefont{Brush}(1966)}]{B66}
\bibinfo{author}{\bibfnamefont{S.~G.} \bibnamefont{Brush}},
  \emph{\bibinfo{title}{Kinetic theory. Irreversible Processes}},
  vol.~\bibinfo{volume}{2} (\bibinfo{publisher}{Pergamon, Oxford},
  \bibinfo{year}{1966}).

\bibitem[{\citenamefont{Chapman and Cowling}(1970)}]{CC70}
\bibinfo{author}{\bibfnamefont{C.}~\bibnamefont{Chapman}} \bibnamefont{and}
  \bibinfo{author}{\bibfnamefont{T.~G.} \bibnamefont{Cowling}},
  \emph{\bibinfo{title}{The Mathematical Theory of Non-Uniform Gases}}
  (\bibinfo{publisher}{Cambridge University Press, Cambridge},
  \bibinfo{year}{1970}), \bibinfo{edition}{3rd} ed.

\bibitem[{\citenamefont{Burnett}(1934)}]{B34}
\bibinfo{author}{\bibfnamefont{D.}~\bibnamefont{Burnett}},
  \bibinfo{journal}{Proc. London Math. Soc.} \textbf{\bibinfo{volume}{40}},
  \bibinfo{pages}{382} (\bibinfo{year}{1934}).

\bibitem[{\citenamefont{Haff}(1983)}]{H83}
\bibinfo{author}{\bibfnamefont{P.~K.} \bibnamefont{Haff}}, \bibinfo{journal}{J.
  Fluid Mech.} \textbf{\bibinfo{volume}{134}}, \bibinfo{pages}{401}
  (\bibinfo{year}{1983}).

\bibitem[{\citenamefont{Jaeger et~al.}(1996)\citenamefont{Jaeger, Nagel, and
  Behringer}}]{JNB96a}
\bibinfo{author}{\bibfnamefont{H.~M.} \bibnamefont{Jaeger}},
  \bibinfo{author}{\bibfnamefont{S.~R.} \bibnamefont{Nagel}}, \bibnamefont{and}
  \bibinfo{author}{\bibfnamefont{R.}~\bibnamefont{Behringer}},
  \bibinfo{journal}{Phys. Today} \textbf{\bibinfo{volume}{49}},
  \bibinfo{pages}{32} (\bibinfo{year}{1996}).

\bibitem[{\citenamefont{Brilliantov and P\"oschel}(2004)}]{BP04}
\bibinfo{author}{\bibfnamefont{N.~S.} \bibnamefont{Brilliantov}}
  \bibnamefont{and}
  \bibinfo{author}{\bibfnamefont{T.}~\bibnamefont{P\"oschel}},
  \emph{\bibinfo{title}{Kinetic Theory of Granular Gases}}
  (\bibinfo{publisher}{Oxford U. P., Oxford}, \bibinfo{year}{2004}).

\bibitem[{\citenamefont{Mehta}(1993)}]{M93}
\bibinfo{editor}{\bibfnamefont{A.}~\bibnamefont{Mehta}}, ed.,
  \emph{\bibinfo{title}{Granular Matter: An Interdisciplinary Approach}}
  (\bibinfo{publisher}{Springer-Verlag, Berlin}, \bibinfo{year}{1993}).

\bibitem[{\citenamefont{Aranson and Tsimring}(2006)}]{AT06}
\bibinfo{author}{\bibfnamefont{I.~S.} \bibnamefont{Aranson}} \bibnamefont{and}
  \bibinfo{author}{\bibfnamefont{L.~S.} \bibnamefont{Tsimring}},
  \bibinfo{journal}{Rev. Mod. Phys.} \textbf{\bibinfo{volume}{78}},
  \bibinfo{pages}{641} (\bibinfo{year}{2006}).

\bibitem[{\citenamefont{Umbanhowar et~al.}(1996)\citenamefont{Umbanhowar, Melo,
  and Swinney}}]{UMS96}
\bibinfo{author}{\bibfnamefont{P.~B.} \bibnamefont{Umbanhowar}},
  \bibinfo{author}{\bibfnamefont{F.}~\bibnamefont{Melo}}, \bibnamefont{and}
  \bibinfo{author}{\bibfnamefont{H.~L.} \bibnamefont{Swinney}},
  \bibinfo{journal}{Nature} \textbf{\bibinfo{volume}{382}},
  \bibinfo{pages}{793} (\bibinfo{year}{1996}).

\bibitem[{\citenamefont{Campbell}(1990)}]{C90}
\bibinfo{author}{\bibfnamefont{C.~S.} \bibnamefont{Campbell}},
  \bibinfo{journal}{Annu. Rev. Fluid Mech.} \textbf{\bibinfo{volume}{22}},
  \bibinfo{pages}{57} (\bibinfo{year}{1990}).

\bibitem[{\citenamefont{Goldhirsch}(2003)}]{Go03}
\bibinfo{author}{\bibfnamefont{I.}~\bibnamefont{Goldhirsch}},
  \bibinfo{journal}{Annu. Rev. Fluid Mech.} \textbf{\bibinfo{volume}{35}},
  \bibinfo{pages}{267} (\bibinfo{year}{2003}).

\bibitem[{\citenamefont{Kadanoff}(1999)}]{K99}
\bibinfo{author}{\bibfnamefont{L.~P.} \bibnamefont{Kadanoff}},
  \bibinfo{journal}{Rev. Mod. Phys.} \textbf{\bibinfo{volume}{71}},
  \bibinfo{pages}{435} (\bibinfo{year}{1999}).

\bibitem[{\citenamefont{Ottino and Khakhar}(2000)}]{OK00}
\bibinfo{author}{\bibfnamefont{J.~M.} \bibnamefont{Ottino}} \bibnamefont{and}
  \bibinfo{author}{\bibfnamefont{D.~V.} \bibnamefont{Khakhar}},
  \bibinfo{journal}{Annu. Rev. Fluid Mech.} \textbf{\bibinfo{volume}{32}},
  \bibinfo{pages}{55} (\bibinfo{year}{2000}).

\bibitem[{\citenamefont{Dufty}(2000)}]{D00}
\bibinfo{author}{\bibfnamefont{J.~W.} \bibnamefont{Dufty}},
  \bibinfo{journal}{J. Phys.: Cond. Matt.} \textbf{\bibinfo{volume}{12}},
  \bibinfo{pages}{A47} (\bibinfo{year}{2000}).

\bibitem[{\citenamefont{Kudrolli}(2004)}]{K04}
\bibinfo{author}{\bibfnamefont{A.}~\bibnamefont{Kudrolli}},
  \bibinfo{journal}{Rep. Prog. Phys.} \textbf{\bibinfo{volume}{67}},
  \bibinfo{pages}{209} (\bibinfo{year}{2004}).

\bibitem[{\citenamefont{Jenkins and Mancini}(1989)}]{JM89}
\bibinfo{author}{\bibfnamefont{J.~T.} \bibnamefont{Jenkins}} \bibnamefont{and}
  \bibinfo{author}{\bibfnamefont{F.}~\bibnamefont{Mancini}},
  \bibinfo{journal}{Phys. Fluids A} \textbf{\bibinfo{volume}{1}},
  \bibinfo{pages}{2050} (\bibinfo{year}{1989}).

\bibitem[{\citenamefont{Goldshtein and Shapiro}(1995)}]{GS95}
\bibinfo{author}{\bibfnamefont{A.}~\bibnamefont{Goldshtein}} \bibnamefont{and}
  \bibinfo{author}{\bibfnamefont{M.}~\bibnamefont{Shapiro}},
  \bibinfo{journal}{J. Fluid Mech.} \textbf{\bibinfo{volume}{282}},
  \bibinfo{pages}{75} (\bibinfo{year}{1995}).

\bibitem[{\citenamefont{Brey et~al.}(1998)\citenamefont{Brey, Dufty, Kim, and
  Santos}}]{BDKS98}
\bibinfo{author}{\bibfnamefont{J.~J.} \bibnamefont{Brey}},
  \bibinfo{author}{\bibfnamefont{J.~W.} \bibnamefont{Dufty}},
  \bibinfo{author}{\bibfnamefont{C.~S.} \bibnamefont{Kim}}, \bibnamefont{and}
  \bibinfo{author}{\bibfnamefont{A.}~\bibnamefont{Santos}},
  \bibinfo{journal}{Phys. Rev. E} \textbf{\bibinfo{volume}{58}},
  \bibinfo{pages}{4638} (\bibinfo{year}{1998}).

\bibitem[{\citenamefont{Garz\'o and Dufty}(1999)}]{GD99}
\bibinfo{author}{\bibfnamefont{V.}~\bibnamefont{Garz\'o}} \bibnamefont{and}
  \bibinfo{author}{\bibfnamefont{J.~W.} \bibnamefont{Dufty}},
  \bibinfo{journal}{Phys. Rev. E} \textbf{\bibinfo{volume}{59}},
  \bibinfo{pages}{5895} (\bibinfo{year}{1999}).

\bibitem[{\citenamefont{Brey and Cubero}(2001)}]{BC01}
\bibinfo{author}{\bibfnamefont{J.~J.} \bibnamefont{Brey}} \bibnamefont{and}
  \bibinfo{author}{\bibfnamefont{D.}~\bibnamefont{Cubero}}, in
  \emph{\bibinfo{booktitle}{Granular Gases}}, edited by
  \bibinfo{editor}{\bibfnamefont{T.}~\bibnamefont{Poschel}} \bibnamefont{and}
  \bibinfo{editor}{\bibfnamefont{S.}~\bibnamefont{Luding}}
  (\bibinfo{publisher}{Springer-Verlag, Berlin}, \bibinfo{year}{2001}),
  Lectures Notes in Physics, pp. \bibinfo{pages}{59--78}.

\bibitem[{\citenamefont{Garz\'o and Dufty}(2002)}]{GD02}
\bibinfo{author}{\bibfnamefont{V.}~\bibnamefont{Garz\'o}} \bibnamefont{and}
  \bibinfo{author}{\bibfnamefont{J.~W.} \bibnamefont{Dufty}},
  \bibinfo{journal}{Phys. Fluids} \textbf{\bibinfo{volume}{14}},
  \bibinfo{pages}{1476–} (\bibinfo{year}{2002}).

\bibitem[{\citenamefont{Lutsko}(2005)}]{L05}
\bibinfo{author}{\bibfnamefont{J.~F.} \bibnamefont{Lutsko}},
  \bibinfo{journal}{Phys. Rev. E} \textbf{\bibinfo{volume}{72}},
  \bibinfo{pages}{021306} (\bibinfo{year}{2005}).

\bibitem[{\citenamefont{Garz\'o
  et~al.}(2007{\natexlab{a}})\citenamefont{Garz\'o, Dufty, and Hrenya}}]{GDH07}
\bibinfo{author}{\bibfnamefont{V.}~\bibnamefont{Garz\'o}},
  \bibinfo{author}{\bibfnamefont{J.~W.} \bibnamefont{Dufty}}, \bibnamefont{and}
  \bibinfo{author}{\bibfnamefont{C.~M.} \bibnamefont{Hrenya}},
  \bibinfo{journal}{Phys. Rev. E} \textbf{\bibinfo{volume}{76}},
  \bibinfo{pages}{031303} (\bibinfo{year}{2007}{\natexlab{a}}).

\bibitem[{\citenamefont{Garz\'o
  et~al.}(2007{\natexlab{b}})\citenamefont{Garz\'o, Hrenya, and Dufty}}]{GHD07}
\bibinfo{author}{\bibfnamefont{V.}~\bibnamefont{Garz\'o}},
  \bibinfo{author}{\bibfnamefont{C.~M.} \bibnamefont{Hrenya}},
  \bibnamefont{and} \bibinfo{author}{\bibfnamefont{J.~W.} \bibnamefont{Dufty}},
  \bibinfo{journal}{Phys. Rev. E} \textbf{\bibinfo{volume}{76}},
  \bibinfo{pages}{031304} (\bibinfo{year}{2007}{\natexlab{b}}).

\bibitem[{\citenamefont{Santos et~al.}(2004)\citenamefont{Santos, Garz\'o, and
  Dufty}}]{SGD04}
\bibinfo{author}{\bibfnamefont{A.}~\bibnamefont{Santos}},
  \bibinfo{author}{\bibfnamefont{V.}~\bibnamefont{Garz\'o}}, \bibnamefont{and}
  \bibinfo{author}{\bibfnamefont{J.~W.} \bibnamefont{Dufty}},
  \bibinfo{journal}{Phys. Rev. E} \textbf{\bibinfo{volume}{69}},
  \bibinfo{pages}{061303} (\bibinfo{year}{2004}).

\bibitem[{\citenamefont{Bird}(1994)}]{B94}
\bibinfo{author}{\bibfnamefont{G.~I.} \bibnamefont{Bird}},
  \emph{\bibinfo{title}{Molecular Gas Dynamics and the Direct Simulation of Gas
  Flows}} (\bibinfo{publisher}{Clarendon, Oxford}, \bibinfo{year}{1994}).

\bibitem[{\citenamefont{Rapaport}(2004)}]{R04}
\bibinfo{author}{\bibfnamefont{D.~C.} \bibnamefont{Rapaport}},
  \emph{\bibinfo{title}{The Art of Molecular Dynamics Simulations}}
  (\bibinfo{publisher}{Cambridge University Press, Cambridge},
  \bibinfo{year}{2004}), \bibinfo{edition}{2nd} ed.

\bibitem[{\citenamefont{Garz\'o and Santos}(2003)}]{GS03}
\bibinfo{author}{\bibfnamefont{V.}~\bibnamefont{Garz\'o}} \bibnamefont{and}
  \bibinfo{author}{\bibfnamefont{A.}~\bibnamefont{Santos}},
  \emph{\bibinfo{title}{Kinetic Theory of Gases in Shear Flows. Nonlinear
  Transport}} (\bibinfo{publisher}{Kluwer Academic Publishers, Dordrecht},
  \bibinfo{year}{2003}).

\bibitem[{\citenamefont{Brey et~al.}(1997)\citenamefont{Brey, Ruiz-Montero, and
  Moreno}}]{BRM97}
\bibinfo{author}{\bibfnamefont{J.~J.} \bibnamefont{Brey}},
  \bibinfo{author}{\bibfnamefont{M.~J.} \bibnamefont{Ruiz-Montero}},
  \bibnamefont{and} \bibinfo{author}{\bibfnamefont{F.}~\bibnamefont{Moreno}},
  \bibinfo{journal}{Phys. Rev. E} \textbf{\bibinfo{volume}{55}},
  \bibinfo{pages}{2846} (\bibinfo{year}{1997}).

\bibitem[{\citenamefont{Tij et~al.}(2001)\citenamefont{Tij, Tahiri, Montanero,
  Garz\'o, Santos, and Dufty}}]{TTMGSD01}
\bibinfo{author}{\bibfnamefont{M.}~\bibnamefont{Tij}},
  \bibinfo{author}{\bibfnamefont{E.~E.} \bibnamefont{Tahiri}},
  \bibinfo{author}{\bibfnamefont{J.~M.} \bibnamefont{Montanero}},
  \bibinfo{author}{\bibfnamefont{V.}~\bibnamefont{Garz\'o}},
  \bibinfo{author}{\bibfnamefont{A.}~\bibnamefont{Santos}}, \bibnamefont{and}
  \bibinfo{author}{\bibfnamefont{J.~W.} \bibnamefont{Dufty}},
  \bibinfo{journal}{J. Stat. Phys.} \textbf{\bibinfo{volume}{103}},
  \bibinfo{pages}{1035} (\bibinfo{year}{2001}).

\bibitem[{\citenamefont{Lutsko}(2006)}]{L06}
\bibinfo{author}{\bibfnamefont{J.~F.} \bibnamefont{Lutsko}},
  \bibinfo{journal}{Phys. Rev. E} \textbf{\bibinfo{volume}{73}},
  \bibinfo{pages}{021302} (\bibinfo{year}{2006}).

\bibitem[{\citenamefont{Garz\'o}(2006)}]{G06}
\bibinfo{author}{\bibfnamefont{V.}~\bibnamefont{Garz\'o}},
  \bibinfo{journal}{Phys. Rev. E} \textbf{\bibinfo{volume}{73}},
  \bibinfo{pages}{021304} (\bibinfo{year}{2006}).

\bibitem[{\citenamefont{{Vega Reyes} and Urbach}(2009)}]{VU09}
\bibinfo{author}{\bibfnamefont{F.}~\bibnamefont{{Vega Reyes}}}
  \bibnamefont{and} \bibinfo{author}{\bibfnamefont{J.~S.}
  \bibnamefont{Urbach}}, \bibinfo{journal}{J. Fluid Mech.}
  \textbf{\bibinfo{volume}{636}}, \bibinfo{pages}{279} (\bibinfo{year}{2009}).

\bibitem[{\citenamefont{Santos et~al.}(2009)\citenamefont{Santos, Garz\'o, and
  {Vega Reyes}}}]{SGV09}
\bibinfo{author}{\bibfnamefont{A.}~\bibnamefont{Santos}},
  \bibinfo{author}{\bibfnamefont{V.}~\bibnamefont{Garz\'o}}, \bibnamefont{and}
  \bibinfo{author}{\bibfnamefont{F.}~\bibnamefont{{Vega Reyes}}},
  \bibinfo{journal}{Eur. Phys. J. Special Topics}
  \textbf{\bibinfo{volume}{179}}, \bibinfo{pages}{141} (\bibinfo{year}{2009}).

\bibitem[{\citenamefont{{Vega Reyes} et~al.}(2010)\citenamefont{{Vega Reyes},
  Santos, and Garz\'o}}]{VSG10}
\bibinfo{author}{\bibfnamefont{F.}~\bibnamefont{{Vega Reyes}}},
  \bibinfo{author}{\bibfnamefont{A.}~\bibnamefont{Santos}}, \bibnamefont{and}
  \bibinfo{author}{\bibfnamefont{V.}~\bibnamefont{Garz\'o}},
  \bibinfo{journal}{Phys. Rev. Lett.} \textbf{\bibinfo{volume}{104}},
  \bibinfo{pages}{028001} (\bibinfo{year}{2010}).

\bibitem[{\citenamefont{Brey et~al.}(1999)\citenamefont{Brey, Dufty, and
  Santos}}]{BDS99}
\bibinfo{author}{\bibfnamefont{J.~J.} \bibnamefont{Brey}},
  \bibinfo{author}{\bibfnamefont{J.~W.} \bibnamefont{Dufty}}, \bibnamefont{and}
  \bibinfo{author}{\bibfnamefont{A.}~\bibnamefont{Santos}},
  \bibinfo{journal}{J. Stat. Phys.} \textbf{\bibinfo{volume}{97}},
  \bibinfo{pages}{281} (\bibinfo{year}{1999}).

\bibitem[{\citenamefont{Montanero et~al.}(1994)\citenamefont{Montanero, Alaoui,
  Santos, and Garz\'o}}]{MMSG94}
\bibinfo{author}{\bibfnamefont{J.~M.} \bibnamefont{Montanero}},
  \bibinfo{author}{\bibfnamefont{M.}~\bibnamefont{Alaoui}},
  \bibinfo{author}{\bibfnamefont{A.}~\bibnamefont{Santos}}, \bibnamefont{and}
  \bibinfo{author}{\bibfnamefont{V.}~\bibnamefont{Garz\'o}},
  \bibinfo{journal}{Phys. Rev. E} \textbf{\bibinfo{volume}{49}},
  \bibinfo{pages}{367} (\bibinfo{year}{1994}).

\bibitem[{\citenamefont{Tritton}(1988)}]{T88}
\bibinfo{author}{\bibfnamefont{J.}~\bibnamefont{Tritton}},
  \emph{\bibinfo{title}{Physical Fluid Dynamics}} (\bibinfo{publisher}{Oxford
  U. P., Oxford}, \bibinfo{year}{1988}).

\bibitem[{\citenamefont{Grad}(1949)}]{G49}
\bibinfo{author}{\bibfnamefont{H.}~\bibnamefont{Grad}},
  \bibinfo{journal}{Commun. Pure Appl. Math.} \textbf{\bibinfo{volume}{2}},
  \bibinfo{pages}{331} (\bibinfo{year}{1949}).

\bibitem[{\citenamefont{Herdegen and Hess}(1982)}]{HH82}
\bibinfo{author}{\bibfnamefont{N.}~\bibnamefont{Herdegen}} \bibnamefont{and}
  \bibinfo{author}{\bibfnamefont{S.}~\bibnamefont{Hess}},
  \bibinfo{journal}{Physica A} \textbf{\bibinfo{volume}{115}},
  \bibinfo{pages}{281} (\bibinfo{year}{1982}).

\bibitem[{\citenamefont{Tsao and Koch}(1995)}]{TK95}
\bibinfo{author}{\bibfnamefont{H.-K.} \bibnamefont{Tsao}} \bibnamefont{and}
  \bibinfo{author}{\bibfnamefont{D.~L.} \bibnamefont{Koch}},
  \bibinfo{journal}{J. Fluid Mech.} \textbf{\bibinfo{volume}{296}},
  \bibinfo{pages}{211} (\bibinfo{year}{1995}).

\bibitem[{\citenamefont{Santos and Astillero}(2005)}]{SA05}
\bibinfo{author}{\bibfnamefont{A.}~\bibnamefont{Santos}} \bibnamefont{and}
  \bibinfo{author}{\bibfnamefont{A.}~\bibnamefont{Astillero}},
  \bibinfo{journal}{Phys. Rev. E} \textbf{\bibinfo{volume}{72}},
  \bibinfo{pages}{031308} (\bibinfo{year}{2005}).

\bibitem[{\citenamefont{{Vega Reyes} et~al.}(2007)\citenamefont{{Vega Reyes},
  Garz\'o, and Santos}}]{VGS07}
\bibinfo{author}{\bibfnamefont{F.}~\bibnamefont{{Vega Reyes}}},
  \bibinfo{author}{\bibfnamefont{V.}~\bibnamefont{Garz\'o}}, \bibnamefont{and}
  \bibinfo{author}{\bibfnamefont{A.}~\bibnamefont{Santos}},
  \bibinfo{journal}{Phys. Rev. E} \textbf{\bibinfo{volume}{75}},
  \bibinfo{pages}{061306} (\bibinfo{year}{2007}).

\bibitem[{\citenamefont{Lobkovsky et~al.}(2009)\citenamefont{Lobkovsky, {Vega
  Reyes}, and Urbach}}]{LVU09}
\bibinfo{author}{\bibfnamefont{A.~E.} \bibnamefont{Lobkovsky}},
  \bibinfo{author}{\bibfnamefont{F.}~\bibnamefont{{Vega Reyes}}},
  \bibnamefont{and} \bibinfo{author}{\bibfnamefont{J.~S.}
  \bibnamefont{Urbach}}, \bibinfo{journal}{Eur. Phys. J. Special Topics}
  \textbf{\bibinfo{volume}{179}}, \bibinfo{pages}{113} (\bibinfo{year}{2009}).

\bibitem[{\citenamefont{{Vega Reyes} et~al.}(2008)\citenamefont{{Vega Reyes},
  Garz\'o, and Santos}}]{VGS08}
\bibinfo{author}{\bibfnamefont{F.}~\bibnamefont{{Vega Reyes}}},
  \bibinfo{author}{\bibfnamefont{V.}~\bibnamefont{Garz\'o}}, \bibnamefont{and}
  \bibinfo{author}{\bibfnamefont{A.}~\bibnamefont{Santos}},
  \bibinfo{journal}{J. Stat. Mech.} \textbf{\bibinfo{volume}{P09003}}
  (\bibinfo{year}{2008}).

\bibitem[{\citenamefont{Hadjiconstantinou
  et~al.}(2003)\citenamefont{Hadjiconstantinou, Garcia, Bazant, and
  He}}]{HGBH03}
\bibinfo{author}{\bibfnamefont{N.}~\bibnamefont{Hadjiconstantinou}},
  \bibinfo{author}{\bibfnamefont{A.~L.} \bibnamefont{Garcia}},
  \bibinfo{author}{\bibfnamefont{M.~Z.} \bibnamefont{Bazant}},
  \bibnamefont{and} \bibinfo{author}{\bibfnamefont{G.}~\bibnamefont{He}},
  \bibinfo{journal}{J. Comput. Phys.} \textbf{\bibinfo{volume}{187}},
  \bibinfo{pages}{274} (\bibinfo{year}{2003}).

\bibitem[{\citenamefont{Astillero and Santos}(2005)}]{AS05}
\bibinfo{author}{\bibfnamefont{A.}~\bibnamefont{Astillero}} \bibnamefont{and}
  \bibinfo{author}{\bibfnamefont{A.}~\bibnamefont{Santos}},
  \bibinfo{journal}{Phys. Rev. E} \textbf{\bibinfo{volume}{72}},
  \bibinfo{pages}{031309} (\bibinfo{year}{2005}).

\bibitem[{\citenamefont{Montanero and Garz\'o}(2002)}]{MG02}
\bibinfo{author}{\bibfnamefont{J.~M.} \bibnamefont{Montanero}}
  \bibnamefont{and} \bibinfo{author}{\bibfnamefont{V.}~\bibnamefont{Garz\'o}},
  \bibinfo{journal}{Physica A} \textbf{\bibinfo{volume}{310}},
  \bibinfo{pages}{17} (\bibinfo{year}{2002}).

\bibitem[{\citenamefont{Lutsko}(2004)}]{L04}
\bibinfo{author}{\bibfnamefont{J.~F.} \bibnamefont{Lutsko}},
  \bibinfo{journal}{Phys. Rev. E} \textbf{\bibinfo{volume}{70}},
  \bibinfo{pages}{061101} (\bibinfo{year}{2004}).

\end{thebibliography}

\end{document}